\documentclass[journal]{IEEEtran}
\usepackage{multirow}
\usepackage{subfigure}
\usepackage{graphicx}
\usepackage{dblfloatfix}
\usepackage{booktabs}
\usepackage{url}
\usepackage{gensymb}

\begin{document}

\title{Defining Gaze Patterns for Process Model Literacy - Exploring Visual Routines in Process Models with Diverse Mappings}

\author{Michael Winter, Heiko Neumann, R\"udiger Pryss, Thomas Probst, and Manfred Reichert
\thanks{Michael Winter and Manfred Reichert are with the Institute of Databases and Information Systems, Ulm University, Ulm, Germany;
(e-mail: michael.winter@uni-ulm.de, manfred.reichert@uni-ulm.de).}
\thanks{Heiko Neumann is with the Institute of Neural Information Processing, Ulm University, Ulm, Germany; (e-mail: heiko.neumann@uni-ulm.de).}
\thanks{R\"udiger Pryss is with the Institute of Clinical Epidemiology and Biometry, University of W\"urzburg, W\"urzburg, Germany; (e-mail: ruediger.pryss@uni-wuerzburg.de).}
\thanks{Thomas Probst is with the Department for Psychotherapy and Biopsychosocial Health, Danube University Krems, Krems, Austria; (e-mail: thomas.probst@donau-uni.ac.at).}}

\markboth{Winter \textit{\MakeLowercase{et al.}:}  Gaze Patterns for Process Model Literacy}%
{Winter \MakeLowercase{\text it{et al.}}: Gaze Patterns for Process Model Literacy}

\maketitle

\begin{abstract}
Process models depict crucial artifacts for organizations regarding documentation, communication, and collaboration. The proper comprehension of such models is essential for an effective application. An important aspect in process model literacy constitutes the question how the information presented in process models is extracted and processed by the human visual system? For such visuospatial tasks, the visual system deploys a set of elemental operations, from whose compositions different visual routines are produced. This paper provides insights from an exploratory eye tracking study, in which visual routines during process model comprehension were contemplated. More specifically, n = 29 participants were asked to comprehend n = 18 process models expressed in the Business Process Model and Notation 2.0 reflecting diverse mappings (i.e., straight, upward, downward) and complexity levels. The performance measures indicated that even less complex process models pose a challenge regarding their comprehension. The upward mapping confronted participants' attention with more challenges, whereas the downward mapping was comprehended more effectively. Based on recorded eye movements, three gaze patterns applied during model comprehension were derived. Thereupon, we defined a general model which identifies visual routines and corresponding elemental operations during process model comprehension. Finally, implications for practice as well as research and directions for future work are discussed in this paper. 
\end{abstract}

\IEEEpeerreviewmaketitle

\section{Introduction}
\label{intro}

A process model constitutes a pictorial graph of information (e.g., data, resources) relying on notational-specific visualization techniques \cite{dumas2013fundamentals}. Process models allow for the creation of abstractions from the physical world, with the application of a variety of symbolic representations in order to delineate the execution of algorithms, the functionality of complex technological systems, or for the standardization of involved processes in organizations. Regarding the latter, process models are applied across different domains (e.g., healthcare \cite{ahmed2019business}, business \cite{rizun2021analyzing}, industry \cite{yin2016service}) for the intention of appropriate process documentation, information communication, and collaboration. Moreover, process models provide additional means facilitating process optimization, problem solving, and decision making \cite{vom2016role}. For this reason, particular attention should be paid that these models are being presented in such a way that the comprehension of information therein does not pose any difficulties. 

In this context, process model literacy is concerned with the reading and comprehension of information presented in process models (i.e., process model comprehension). Process model literacy is a specialization of graph literacy depicting the capability to extract as well as to comprehend graphically presented information and to draw inferences based thereof \cite{galesic2011graph}. Due to the relevance and widespread application of process models in different domains, still many unresolved issues regarding the factors thwarting the comprehension of such models exist and, as a consequence, the identification of these factors becomes crucial \cite{dikici2018factors}. In addition, unraveled factors as well as best practices benefiting process model literacy should also continue to be the subject of research (e.g., for potential further improvements). On that account, a vast body of research was created in the past on studying the factors affecting process model literacy. 

Thereby, two main branches of research have emerged in the context of process model literacy: the study of \raisebox{.5pt}{\textcircled{\raisebox{-.9pt} {1}}} objective (i.e., process model properties) and \raisebox{.5pt}{\textcircled{\raisebox{-.9pt} {2}}} subjective (i.e., person-related characteristics) factors and their influence on the reading and comprehension of process models \cite{figl2017comprehension}. Regarding \raisebox{.5pt}{\textcircled{\raisebox{-.9pt} {1}}} objective factors, research addressed various structural aspects in a process model \cite{dikici2018factors}. For example, the works in \cite{jovst2016empirical}, \cite{haisjackl2016understanding}, and \cite{kluza2017comparison} conducted a comparison across different process modeling notations and their influence on process model comprehension. In turn, the authors in \cite{koschmider2016comprehensive} and \cite{corradini2018guidelines} propose directives for better comprehensible process models. Further, effects and guidelines for proper labeling of modeling elements are discussed in \cite{leopold2015learning} as well as in \cite{van2016detecting}. Moreover, process model quality (i.e., semantic, syntactic, pragmatic) as a supportive factor for process model literacy in general is discussed in \cite{krogstie2016quality}. Additionally, a comprehensive overview of existing metrics for the measurement of process model quality is discussed in \cite{kahloun2016quality}. Finally, the works from \cite{schoknecht2017similarity} and \cite{van2016business} discuss metrics allowing the measurement of compliance with directives that foster process model comprehension. 

Regarding \raisebox{.5pt}{\textcircled{\raisebox{-.9pt} {2}}} subjective factors, character traits (see \cite{recker2010continued}) of a model reader in the context of process model literacy have been investigated. These include, among others, general knowledge about conceptual modeling as an integral aspect in process model literacy \cite{de2015systematic}. In this context, specific expertise in working with process models is another crucial success factor facilitating the comprehension of such models \cite{gassen2015towards}. Further, research in \cite{reijers2010study} depicts professional background as well as domain familiarity as additional contributors towards a proper comprehension of process models. In turn, the authors from the work presented in \cite{mendling2010activity} address the effects of domain familiarity on model comprehension.

In the recent past, as can be seen in other domains (e.g., \cite{mohanani2018cognitive}), subjective factors are considered in a broader spectrum, with a particular emphasis on cognitive aspects. For example, the authors in \cite{turetken2017cognitive} investigate cognitive style (i.e., thinking style) during the comprehension of process models. Moreover, the effects of different cognitive abilities and learning styles with respect to process model literacy are presented in \cite{recker2014process}. Finally, the effects of cognitive biases during process model comprehension are shown in \cite{razavian2016cognitive}. 

In this context, the proliferation of modern sensing technologies (e.g., serious games \cite{winter2020learning}, mobile devices \cite{tallon2019comprehension}, electrodermal activity \cite{winter2020towards}) enabled a more fine-grained investigation of cognitive aspects in process model literacy. Thereby, the application of eye tracking constitutes a prominent methodology in order to obtain novel insights in process model comprehension.

In \cite{zimoch2017eye}, experiences and lessons learned about the usage of eye tracking in the context of process model literacy are discussed. Further, as shown by the authors in \cite{bera2019using}, different eye movements parameters may constitute indicators for high cognitive load, whereas research in \cite{petrusel2017visual} and \cite{winter2021applying} demonstrated that visual cues support the reading and comprehension of process models. Finally, the authors in \cite{petrusel2017visual} provide insights into how visual cognition affects process model comprehension. 

However, although existing research addressed how process models are read and comprehended by model readers (e.g., \cite{mendling2019empirical,haisjackl2018humans}), it is still not clear, especially from a cognitive point of view, how information presented in process models, concerning the syntactic, pragmatic, and semantic quality dimension, is captured, processed, and comprehended by the human visual system. More precisely, does a systematic mechanism influencing our focus of attention in process model comprehension exist, thus resulting in common as well as different comprehension strategies? In this context, the author in \cite{ullman1987visual} proposed a versatile visual analysis approach, which nature constitutes visual routines. A visual routine describes the composition of elemental operations (e.g., anchoring \cite{pylyshyn1989role}) from the visual system, which are deployed during visuospatial tasks. Based on this approach, related research contributed additional elemental operations contributing to further visual routines deployed by the visual system for the extraction of information from visual scenes \cite{foulsham2015eye, geyer2007eye}. Thereupon the question arises, which elemental operations are or can be relevant in the context of process model comprehension and how are these operations combined to visual routines?

In order to investigate the latter, this paper provides the first results from an exploratory eye tracking study, in which visual routines and corresponding elemental operations in process model comprehension were evaluated. Based on the results obtained, specific gaze patterns could be identified that summarize common visual routines during process model comprehension. Furthermore, we define a model that elaborates how different objective-related factors during the comprehension of process models evoke the deployment of visual routines, which are composed of elemental operations. In general, to the best of our knowledge, there exist no works dealing with visual routines in the context of process model literacy so far. For this reason, the insights presented in this paper shall set the empirical foundation for further studies that will evaluate visual routines in process model comprehension to foster our general understanding of working with such models.

The structure of this paper is as follows: Section II provides theoretical background about visual routines and elemental operations. Materials and methods of the exploratory study are described in Section III. In Section IV, obtained results are presented descriptively and tested for significance. A discussion about further eye movements analyses (i.e., gaze patterns) and the defined model regarding visual routine deployment in the context of process model comprehension are presented in this section. Moreover, Section IV provides implications, limiting factors and gives an outlook on future work. Finally, Section V summarizes the paper. 

\section{Theoretical Background}
\label{theo}

The human visual system offers a wide range of flexible spatial analysis mechanisms for graph comprehension \cite{lai2016measuring}. For example, the visual system is able, even without conscious effort, to process quantitative information such as x increases exponentially as y decreases in a line graph as well as the spatial relation between graph elements (e.g., identification of the largest area in a pie chart). Thereby, cognitive research showed that three major factors exist, which influence the interpretation of graphs: \raisebox{.5pt}{\textcircled{\raisebox{-.9pt} {1}}} knowledge in graph literacy \cite{okan2016people}, \raisebox{.5pt}{\textcircled{\raisebox{-.9pt} {2}}} expectations about the data in a graph \cite{shah2011bar}, and \raisebox{.5pt}{\textcircled{\raisebox{-.9pt} {3}}} visual properties of a graph \cite{okan2012higher}. The work at hand puts an emphasis on the latter factor \raisebox{.5pt}{\textcircled{\raisebox{-.9pt} {3}}}. In this context, related studies demonstrated that the processing of visual representations is bound to limitations in the visual system (e.g., the processing of only a few elements and respective properties is possible \cite{ceja2020capacity}). Furthermore, it was shown that the efficiency in the comprehension of graphs is constrained by the sequence single elements in a graph are considered \cite{michal2017visual}. In more detail, specific orientation points in a graph (i.e., anchors) are decisive, which determine further processing sequences. In order to cope with these visual capacity limitations, the visual system needs to deploy a systematic approach for the proper processing of information in graphs. In this context, the author in \cite{ullman1987visual} proposed a conceptual framework describing that the visual spatial analysis evokes a set of elemental visual operations, which analyze and define spatial properties of single elements (e.g., shape) as well as the relation of multiple elements in a graph (e.g., topology). Further, these elemental operations are combined with each other and, thus, result in so-called visual routines. Thereby, different visual inputs or objectives result in various visual routines during visuospatial information processing. For visual routines, \cite{ullman1987visual} defined the following five elemental operations:

\begin{itemize}
	\item Shifting of the processing focus: Application of elemental operations in different locations in a graph (e.g., information extraction).
	\item Indexing of salient elements: Odd-man-out representations (e.g., indexable locations such as a yellow dot between green ones on a blue area) that attract attention and represent anchors for further processing. 
	\item Bounded activation or coloring: The separation and individual consideration of visual units from the composite of a monolithic structure (e.g., inside/outside relation).
	\item Boundary tracing: Mental spreading of the processing focus along boundaries and contours.
	\item Marking: Holding on to a visual structure, unit, or location as a congruence point that can be referred to in the future or which is not relevant for further processing and, therefore, can be ignored.
\end{itemize}

As a result, the combination of the elemental operations in the visual analysis of graphs produces a variety of independent visual routines. Vice versa, a visual routine is composed of a set of elemental operations. Thereby, the information captured in a graph is extracted, processed, and comprehended in a visual routine. The proposed elemental operations can run in parallel or are applied sequentially on a specific location in a graph. According to \cite{ullman1987visual}, visual analysis in graph literacy is described as an incremental process, in which the produced visual routines are composed together, towards a big picture. As a result, visual routines and associated elemental operations allow the visual system to handle sophisticated visuospatial tasks (e.g., count the number of all green circles). Thereby, research suggests that the deployment of visual routines is controlled by the cognitive control (i.e., responsible for attention shift) \cite{michal2016visual}. The latter comprises mental functions that are responsible for human behavior (i.e., the focus of attention) in order to achieve a specific objective (e.g., count all red balloons) \cite{botvinick2015motivation}. Thereby, the visual system is part of the cognitive control \cite{babin2006executive}.

Based on the proposed conceptual framework, related work investigated the application of visual routines in graph comprehension, revealing additional operations and visual routines. For example, the study of attention and visual search is the subject of the work presented in \cite{wolfe2015} demonstrating that different operations are executed in order to find specific information. \cite{yang2002visual} addresses visual search and cueing providing evidence that the search for objects may be executed in parallel or sequential. Further, visual association (i.e., observable vs. non-observable) as an important operation that utilizes contextual information to foster the understanding in visual scenes is discussed in \cite{torralba2006contextual}. Moreover, forward planning constitutes a visual routine deployed in everyday tasks, in which objects for future application are identified and the motor system is activated to carry out these tasks \cite{land2001ways}. The work presented in \cite{tan2017vision} proposes a hierarchical framework comprised of visual routines (e.g., saccade planning, visual buffer) that is utilized during visual tasks in the real world. The authors in \cite{ballard2009modelling} demonstrate that the deployment of visual routines is not driven by discontinuities (e.g., color) in a visual scene, but objective-based. Finally, some computational image processing approaches are utilizing visual routines \cite{tinoco2018data,hsiao2020role,olague2019brain}.

\begin{table}[!b]
	\centering
	\caption{Sample Description and Demographics}
	\label{Sample}
	\begin{tabular}{l|c|c|c|c|c}
		ID & Gender & Age & Education & Profession & Expertise \\ \hline \hline
		1  & 1      & 21  & 0         & 0          & 0 (-)     \\
		2  & 0      & 30  & 2         & 1          & 1 (9)     \\
		3  & 1      & 28  & 2         & 1          & 1 (2)     \\
		4  & 1      & 36  & 3         & 1          & 1 (5)     \\
		5  & 1      & 29  & 2         & 2          & 0 (-)     \\
		6  & 1      & 26  & 2         & 0          & 0 (-)     \\
		7  & 1      & 24  & 2         & 3          & 0 (-)     \\
		8  & 1      & 43  & 3         & 1          & 1 (4)     \\
		9  & 1      & 46  & 3         & 0          & 0 (-)     \\
		10 & 1      & 32  & 3         & 0          & 0 (-)     \\
		11 & 1      & 25  & 3         & 0          & 0 (-)     \\
		12 & 0      & 29  & 2         & 4          & 1 (1)     \\
		13 & 0      & 33  & 2         & 4          & 0 (-)     \\
		14 & 1      & 35  & 3         & 0          & 0 (-)     \\
		15 & 0      & 31  & 2         & 4          & 0 (-)     \\
		16 & 1      & 28  & 2         & 1          & 1 (3)     \\
		17 & 1      & 28  & 2         & 1          & 1 (4)     \\
		18 & 0      & 27  & 2         & 1          & 1 (4)     \\
		19 & 0      & 23  & 0         & 1          & 1 (3)     \\
		20 & 0      & 30  & 2         & 1          & 1 (3)     \\
		21 & 0      & 34  & 2         & 5          & 1 (3)     \\
		22 & 0      & 24  & 1         & 1          & 1 (3)     \\
		23 & 1      & 30  & 2         & 1          & 1 (4)     \\
		24 & 1      & 35  & 2         & 1          & 0 (-)     \\
		25 & 1      & 25  & 2         & 1          & 1 (3)     \\
		26 & 1      & 28  & 2         & 1          & 1 (2)     \\
		27 & 1      & 60  & 2         & 5          & 0 (-)     \\
		28 & 0      & 30  & 2         & 6          & 0 (-)     \\
		29 & 1      & 58  & 1         & 5          & 0 (-)     \\ \hline \hline
		\multicolumn{6}{l}{\textbf{Note:} Gender: 0 = male, 1 = female; Education: 0 = abitur, 1 =} \\  
		\multicolumn{6}{l}{bachelor's degree, 2 = master's degree, 3 = doctor of philosophy} \\    
		\multicolumn{6}{l}{Profession: 0 = medicine, 1 = computer science, 2 = economics,} \\ 
		\multicolumn{6}{l}{3 = mathematics, 4 = epidemiology, 5 = pedagogy; Expertise: 0 = } \\    
		\multicolumn{6}{l}{No, 1 = Yes (score in knowledge test; max. was 10)}       
	\end{tabular}
\end{table}

\begin{figure*}[t]
	\centering
	\includegraphics[width=\linewidth]{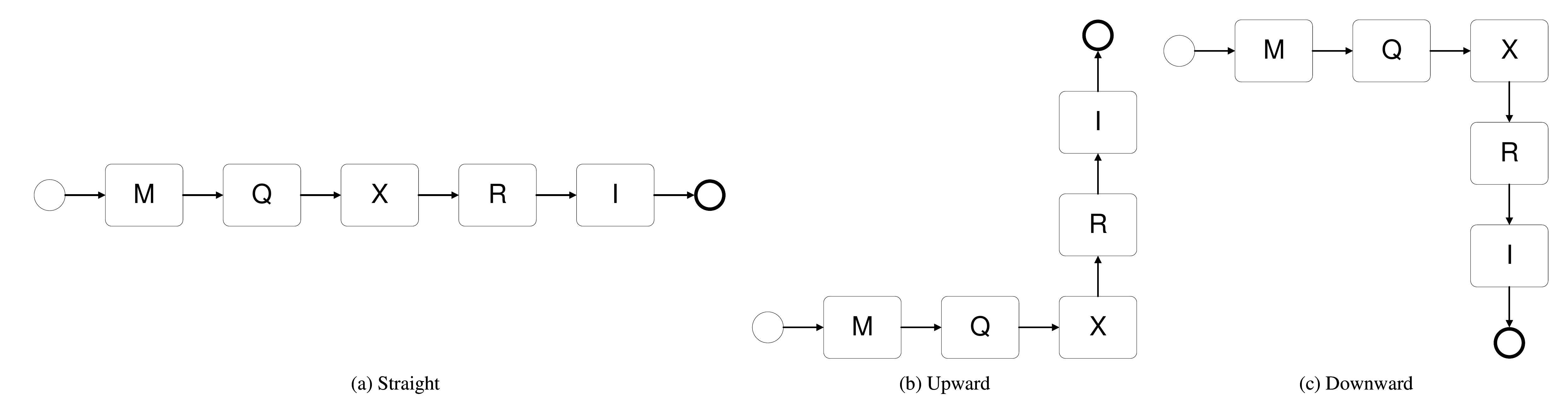}
	\caption{Medium difficulty process models with diverse mapping showing respective start and end events, five activities with a non lexicographic labeling, and connecting sequence flows: (a) straight, (b) upward, and (c) downward}
	\label{model}
\end{figure*}

\section{Materials and Methods}
\label{method}

\subsection{Study Context}
\label{context}

Process models facilitate the exchange of important information and are used - inter alia - for documentation, decision making, and problem solving as well as optimization and automation \cite{weber2011refactoring}. Thereby, the comprehension of these models constitutes a complex task in process model literacy, as contained relevant syntactic, semantic, and pragmatic information in such models needs to be parsed correctly by the visual system \cite{krogstie2016quality}. Since the capacities of the visual system are limited and according to the framework proposed in \cite{ullman1987visual}, visual routines and elemental operations (i.e., of which these routines are composed of; see Section~\ref{theo}) are deployed as a systematic approach for the processing of respective information with an emphasis on visuospatial properties. Existing literature already contributed majorly in the context of process model literacy (see Section~\ref{intro}). However, the deployment of visual routines and associated elemental operations during the reading and comprehension of process models have not been addressed so far in prior work. For this reason, the paper presents the first insights from an exploratory eye tracking study, in which eye movements on process models with diverse mappings (i.e., straight, upward, downward) and complexity levels (i.e., easy, medium, hard) were employed in order to get a better understanding of how the visual system shifts attention and captures information presented in these models. The presentation of the results is as follows: First, the process model mappings as well as complexity levels and related performance measures are juxtaposed to examine whether they showed different characteristics during process model comprehension. Second, based on recorded eye movements, three gaze patterns (i.e., orientation, comprehension, congruence) are presented that we identified during the comprehension of process models. Third, we propose a model that is concerned with the deployment of visual routines and elemental operations during process model comprehension. The findings contribute to the recent body of research towards a cognitive-centric perspective in process model comprehension. Further, the investigation of visual routines (i.e., how the visual system, respective cognitive control, extracts and process information from such models) offers a new perspective enabling a more sophisticated approach to study the factors affecting process model comprehension (i.e., in order to be able to provide appropriate assistance for a model reader) and, hence, will be subject of future work. Finally, the conceptual framework for the comprehension of process models incorporating methods and theories from cognitive neuroscience and psychology presented in \cite{zimoch1} will be enriched based on the insights from this work.

\subsection{Participants}
\label{participants}

The study included a total of n = 29 participants, which were recruited at the University of Würzburg (n = 15; ID 1 - 15) and Ulm University (n = 14; ID 16 - 29). 10 participants were female, 19 male and the mean (m) age was m = 32.00 years (standard deviation (SD) = 9.10). Regarding the highest education qualification, 2 participants hold abitur, 2 a bachelor's degree, 19 a master's degree, and 6 a doctor of philosophy. Further, 6 participants had indicated working in the field of medicine, 14 in computer science, 4 in economics, 1 in mathematics, 3 in epidemiology, and 1 in pedagogy. 14 participants had no prior expertise in working with process models, whereas 15 participants already had expertise. Moreover, the latter group with expertise needed to complete a knowledge test about syntactical rules in process models for a better categorization regarding process model expertise. This test contained a set of 10 true-or-false comprehension questions. On average, the participants reached in the knowledge test m = 3.53 (1.75) points (maximum was 10 points). Finally, Table~\ref{Sample} summarizes the study sample. 

\subsection{Materials}
\label{material}

In the reported study, n = 18 process models expressed in terms of the BPMN 2.0 were used \cite{OMG2010}. In general, the choice to use BPMN 2.0 process models was made for several rationales: First, BPMN 2.0 is the de facto industry standard for the creation of readily comprehensible process models and, second, an ISO/IEC 1950:2013 standard \cite{ISO13586}. In particular, BPMN 2.0 serves as a seamless link between process design (e.g., process documentation) and implementation as well as enactment (e.g., process automation). Third, during the last decade, a vast body of knowledge evolved, which has promoted the widespread application of BPMN 2.0 in practice as well as in research \cite{corradini2018guidelines}. These 18 process models were divided into 3 different mapping groups, containing 6 models each (i.e., straight, downward, upward). The 6 models in each mapping group differed in the level of complexity based on their number of activities. The simplest process model contained 1 activity, while the most difficult model was made up of 11 activities. An increase in the level of complexity was achieved by adding two more activities each (i.e., number of activities: 1, 3, 5, 7, 9, 11). Thereby, the process models containing 1 and 3 activities were classified with an easy, 5 and 7 activities with a medium, and 9 and 11 with a hard level of complexity. Further, each process model contained one start and one end event. While the process models in the straight mapping consisted only of a linear process flow from left to right, the process flow changed upwards or downwards in the models in the other two mapping types. Regarding the latter two, the display direction change in the process flow occurred at the middle element of the activity chain of the respective process model. Further, each process model had a unique start and end event. Elements within a process model were connected with a sequence flow determining the general process flow. For this exploratory study, the three process flow directions (i.e., straight, upward, downward) were chosen, as they constituted the most used directions for the representation of flow directions (i.e.,  generally in conceptual models \cite{gulden2015toward}). Moreover, the process models were kept intentionally simple in order to minimize the influence of incidental effects caused by non-related factors (e.g., advanced modeling elements; see Section~\ref{future}). The labels of the activities contained only alphabetical characters (i.e., non-semantic context). Note that the activities were not labeled in alphabetical order, but randomly (i.e., arbitrarily). The reason was that a lexicographic order might ease the comprehension process. The Figures~\ref{model} (a) - (c) illustrate from each mapping group a respective process model with 5 activities (i.e., medium level of complexity), respective start as well as end event, and connecting sequence flows. Furthermore, for each process model and depending on the level of complexity, 1 - 3 true-or-false comprehension questions were created that needed to be answered after the comprehension of each model (i.e., 1 question for the easy, 2 for the medium, and 3 for the hard process models). The comprehension questions referred to process model syntactics (e.g., activity X needs to be executed before activity Y) as well as semantics (e.g., activity Z is the first activity). Finally, regarding comparability, the process models and the comprehension questions were shared between the three mapping groups\footnote{The materials used in the study are available at: \url{https://tinyurl.com/3uwzh6cy}}.

\subsection{Performance Measures}
\label{measures}

\begin{figure}[t]
	\centering
	\includegraphics[width=.95\linewidth]{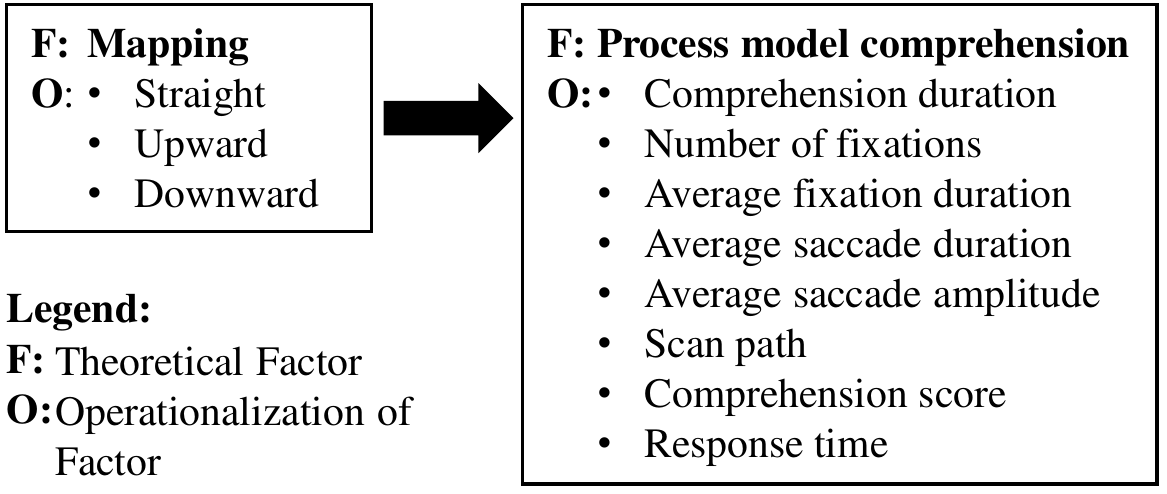}
	\caption{Research model for the defined performance measures}
	\label{rm}
\end{figure}

In the following, the considered performance measures in the study are described in detail. Furthermore, Figure~\ref{rm} summarizes defined measures used for inferential statistics (see Section~\ref{inferential}) in a research model.

\begin{itemize}
\item Comprehension duration: A timestamp was added at the moment participants started comprehending a process model. After comprehending respective model, another timestamp was added indicating the duration. This allowed us to measure the duration needed for comprehension on a fine-grained level.

\item Number of fixations: Fixations constitute eye movements of very low velocity (i.e., eyes remain still on a specific place in a stimulus over a period of time \cite{andersson2017one}). Moreover, fixation sequence analyses (i.e., sequential analysis of attentional content reading) allowed us to, for example, visualize comprehension strategies, make conclusions about performed eye movements in the three mapping types as well as about specific points in the stimulus (i.e., process models) that attracted the attention of the participants in the comprehension process. 

\item Average fixation duration: The fixation duration indicates the period of time in which the eyes remain still while looking at a stimulus \cite{fixi}. During this period of time, the acquisition of information from the currently viewed point in a stimulus (i.e., process model) by the visual system takes place. Hence, the analysis of the average fixation duration allowed for additional assumptions regarding, for example, applied scanning types (e.g., skimming) during the comprehension of the presented process models \cite{korbach2018differentiating}. 

\item Average saccade duration: Usually, a saccade lasts on average 20 - 40 ms and during this time visual reception of information is suppressed. The saccade duration provided additional insights regarding scanning strategies in the comprehension of the process models. 

\item Average saccade amplitude: The amplitude describes the size of the saccade and is measured in degrees. The saccade amplitude is linearly correlated with its duration (i.e., a larger amplitude results in a longer duration and vice versa \cite{unema2005time}). Specific eye movement patterns (e.g., back-and-forth saccade jumps) may be identified with the evaluation of the saccade amplitudes.

\item Scan path: The scan path describes a sequence pattern of shifted attention (i.e., chronological concatenation of fixations and saccades), where information is visually recorded in a specific trajectory, when scanning a stimulus \cite{foulsham2012comparing}. The analysis of the scan paths allowed for the identification of scanning as well as eye movement patterns (e.g., targeted search) during the comprehension of the process models. 

\item Comprehension score: Participants needed to answer for each comprehended process model, dependent on the level of complexity, 1 - 3 true-or-false comprehension questions. The questions referred to the semantic as well as syntactic quality dimensions of the process models. For each correct given answer, a point was awarded. 

\item Response time: The response time for the comprehension questions was measured to investigate whether specific mapping types were more catchy, thus leading to a faster response in answering.

\smallskip

In addition to the defined performance measures, the following information was collected at the end of the study used for analysis in descriptive statistics (see Section~\ref{design})\footnote{The questions regarding cognitive load and level of acceptability are available at: https://tinyurl.com/3uwzh6cy}:

\smallskip

\begin{figure*}[t]
	\centering
	\includegraphics[width=\linewidth]{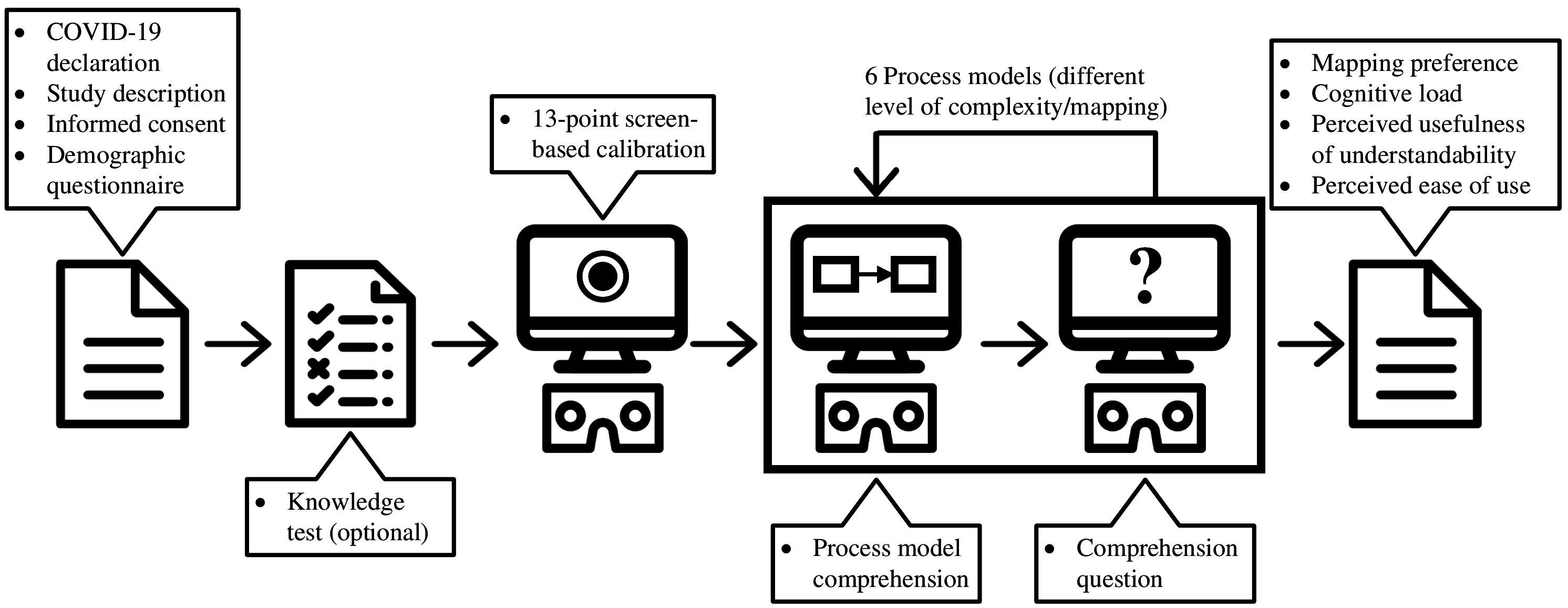}
	\caption{Applied study design}
	\label{fig:design}
\end{figure*}

\item Cognitive load: The cognitive load depicts the invested cognitive capacity of the working memory during a task. The cognitive load consists of the following dimensions: intrinsic, extraneous, and germane cognitive load \cite{sweller1994cognitive}. Intrinsic load describes the complexity of intrinsic information of the presented material and is affected by existing knowledge and element interactivity (e.g., demand on the working memory). Extraneous load, in turn, is influenced by the way information is presented. Finally, germane load delineates the mental effort to process and comprehend information based on constructed mental models \cite{paas2003cognitive}. To measure the single dimensions related to the cognitive load, the adapted measurement discussed in \cite{klepsch2017development} was used in the exploratory study. The latter work demonstrated that the use of this measurement constitutes a validated and reliable instrument for measuring the cognitive load. As a result, the measurement can be applied from an informed (i.e., with prior knowledge) and naïve point of view (i.e., without prior knowledge) about the concept of cognitive load. The single dimensions, which were comprised of several items (i.e., two for intrinsic, three for extraneous, and two for germane cognitive load), were assessed on a psychometrically standardized questionnaire (i.e., 7-point Likert scale from strongly disagree (i.e., 1) to strongly agree (i.e., 7)).

\item Perceived usefulness for understandability (PUU): Derived from the technology acceptance model (TAM) \cite{king2006meta}, PUU describes the perceived usefulness of a particular process model mapping in the context of process model comprehension. Therefore, four items on a 7-point Likert scale from strongly disagree (i.e., 1) to strongly agree (i.e., 7) needed to be answered totaling to a min/max value of (4 x 7). The used measure was evaluated for validity and reliability in prior research \cite{turetken2019influential}. 

\item Perceived ease of understandability (PEU): Derived from TAM, PEU characterizes that the use of a particular process model mapping is associated with less mental effort during model comprehension. Therefore, four items on a 7-point Likert scale from strongly disagree (i.e., 1) to strongly agree (i.e., 7) needed to be answered totaling to a min/max value of (4 x 7). The used measure was evaluated for validity and reliability in prior research \cite{turetken2019influential}. Both, PUU and PEU, represented the general level of acceptability in the study. 

\item Mapping Preference: The most comprehensible process model mapping was inquired from the participants. All three mapping types were juxtaposed and the participants were asked to indicate the best comprehensible mapping type based on subjective preferences. Accordingly, the frequency of the respective mapping types was evaluated.
\end{itemize}

\subsection{Study Design}
\label{design}

The study was conducted in line with the principles of the Declaration of Helsinki. All materials and methods were approved by the Ethics Committee of Ulm University and were carried out in accordance with the approved guidelines (No. 446/20). Before the study reported in this paper, a pilot study with four participants was conducted. The pilot study was used in order to obviate ambiguities as well as misunderstandings in the study design. Moreover, the overall quality of the study material was increased and technical functions (e.g., data collection) had been checked for their proper implementation. Due to the COVID-19 pandemic, the study was conducted in a prepared lab (e.g., room size of 25 m\textsuperscript{2} with sufficient room ventilation) at Ulm University and the University of Würzburg. Moreover, a dedicated study procedure was carried out in compliance with mandatory COVID-19 hygiene regulations (e.g., permanent wearing of a mouth-nose protection, disinfection of all used instruments before and after each participant). At each study session, only one participant was evaluated, and a session took about 20 minutes for completion. Moreover, the design of the study was based on the guidelines proposed in \cite{Wohlin2000}, which provided all fundamentals for studies in the context of information engineering. The study procedure was as follows: In compliance with the hygiene regulations and after a standardized introduction of the participant, she or he needed to sign a COVID-19 declaration. Afterwards, the procedure of the study was explained, questions as well as ambiguities of the participant regarding the study were addressed, and informed consent was obtained. Following this, demographic data (e.g., age, gender) was collected and expertise in working with process models was asked for. In case of a positive answer regarding the latter, the participant needed to complete a process model knowledge test, which contained a set of 10 true-or-false questions. After the completion of these steps, the participant was placed in front of the eye tracking appliance and the device was calibrated for a precise recording of the eye movements. Then, the participant completed a brief tutorial in order to familiarize them with the functionality of the eye tracking appliance. Hereafter, the participant comprehended a total of six process models. Based on a defined permutation table, two each straight, downward and upward process models were evaluated. Moreover, regarding the level of process model complexity, the six process models consisted of two each easy, medium, and hard models. Thereby, the models were presented in a randomized order. In general, the permutation table ensured an equal distribution of the process models (i.e., mapping and level of complexity) among the participants. After the comprehension of each process model, the participant needed to answer, based on the level of process model complexity, a set of true-or-false comprehension questions (i.e., 1 for easy, 2 for medium, and 3 for hard process models; see Section~\ref{material}). After the comprehension of all six process models, questionnaires capturing the mapping preference, cognitive load, perceived usefulness for understandability, and perceived ease of use had to be answered by the participant. Finally, after the opportunity to leave feedback, the study ended. Figure~\ref{fig:design} outlines the used design in the study.

\subsection{Instrumentation}
\label{instru}

COVID-19 declaration, informed consent, demographic data, knowledge test score, information related to the mapping preference, cognitive load as well as the level of acceptability, and feedback was collected using pen-and-paper based report forms (i.e., questionnaires). Eye movements were recorded with the SMI iView X Hi-Speed system (sampling rate of 240 Hz; chin rest). The tracking appliance was placed in front of a 23'' monitor (resolution of 1920x1080, 96 PPI) presenting the process models to the participants. Participants were seated approximately 65 - 70 cm from the monitor. For calibration, a 13-point screen-based calibration was performed. Before each process model, a fixation cross for the duration of 500 ms was shown. The events (i.e., start and end) had a width of .8\textdegree, activities had a width of 2.2\textdegree\ and a height of 1.8\textdegree, and the sequence flow between the model elements was 1.2\textdegree\ wide. Since the process models were presented centered on the monitor, the fixation cross was placed within the process path in the sequence mapping, whereas it was placed above (i.e., upward) or below (i.e., downward) in the other mapping types. The models were constantly within the convex hull. For answering the true-or-false comprehension questions, participants used a keyboard with two predefined keys providing the respective answering options (i.e., 'true' and 'false'). Eye tracking data collected during the study was analyzed and visualized with SMI BeGaze 3.7. Finally, SPSS 25 was used for all statistical analyses.

\section{Results}
\label{results}

This section presents obtained study results. As described in Section~\ref{context}, in our exploration of visual routines in the context of process model literacy, the evaluated model mappings with related complexity levels are juxtaposed in order to examine differences in considered performance measures (see Section~\ref{descriptive}). Afterwards, discussed descriptive results are tested for significance (see Section~\ref{inferential}). Following this, recorded eye movements are evaluated in more detail (see Section~\ref{eye}) to investigate the association of visual routines and specific gaze patterns during process model comprehension (see Section~\ref{visual pattern}). Moreover, discussion as well as implications (see Section~\ref{discussion}), future work (see Section~\ref{future}), and confronted limitations (see Section~\ref{limit}) are discussed in the final part of this section. The following Table~\ref{models} presents a summary of the evaluated process models. More specifically, the table shows, for each mapping (i.e., straight, upward, downward), corresponding process model (i.e., PM; the numbering of the models is used in the further) with related properties: level of complexity (i.e., Complexity), number of activities (i.e., Activity), number of comprehension questions (i.e., Question), and the total number of evaluated models in the study (i.e., No Study).

\begin{table}[b]
	\caption{Evaluated process models in the exploratory study}
	\label{models}
	\begin{tabular}{l|l|l|c|c|c}
		PM & Mapping       & Complexity    & Activity & Question & No Study \\ \hline \hline
		1  & Straight  & Easy   & 1      & 1    & 10   \\
		2  & Straight  & Easy   & 3      & 1    & 10   \\
		3  & Straight  & Medium & 5      & 2    & 10   \\
		4  & Straight  & Medium & 7      & 2    & 9    \\
		5  & Straight  & Hard   & 9      & 3    & 9    \\
		6  & Straight  & Hard   & 11     & 3    & 10   \\ \hline \hline
		7  & Upward   & Easy   & 1      & 1    & 9    \\
		8  & Upward   & Easy   & 3      & 1    & 10   \\
		9  & Upward   & Medium & 5      & 2    & 10   \\
		10 & Upward   & Medium & 7      & 2    & 10   \\
		11 & Upward   & Hard   & 9      & 3    & 10   \\
		12 & Upward   & Hard   & 11     & 3    & 9    \\ \hline \hline
		13 & Downward & Easy   & 1      & 1    & 10   \\
		14 & Downward & Easy   & 3      & 1    & 9    \\
		15 & Downward & Medium & 5      & 2    & 9    \\
		16 & Downward & Medium & 7      & 2    & 10    \\
		17 & Downward & Hard   & 9      & 3    & 10    \\
		18 & Downward & Hard   & 11     & 3    & 10   \\ \hline \hline
	\multicolumn{6}{l}{\textbf{Note:} PM = Process model}     
	\end{tabular}
\end{table}

\subsection{Descriptive Statistics}
\label{descriptive}

Table~\ref{eyedcsr} presents obtained results regarding eye tracking parameters. More specifically, for each process model (i.e., PM) 1 - 18 (see Table~\ref{models}), mean (m) value as well as the standard deviation (SD) for the following measures are shown: duration (in ms), number of fixations (i.e., No Fixation), fixation duration (i.e., in ms; Fix Duration), saccade duration (i.e., in ms; Sacc Duration), saccade amplitude (i.e., in deg; Sacc Amp), length of the scan path (in px). 

\begin{table*}[h]
	\centering
	\scriptsize
	\caption{Descriptive results regarding eye tracking}
\begin{tabular}{lllllllllllll}
	\multicolumn{1}{l|}{PM} & \multicolumn{2}{c|}{Duration (ms)}         & \multicolumn{2}{c|}{No Fixation}      & \multicolumn{2}{c|}{Fix Duration (ms)} & \multicolumn{2}{c|}{Sacc Duration (ms)} & \multicolumn{2}{l|}{Sacc Amp (deg)} & \multicolumn{2}{c}{Scan path (px)} \\ \hline \hline
	\multicolumn{1}{l|}{1}                                                 & 5286.30  & \multicolumn{1}{l|}{(3215.94)}  & 13.60  & \multicolumn{1}{l|}{(6.09)}  & 252.46  & \multicolumn{1}{l|}{(62.27)} & 40.15   & \multicolumn{1}{l|}{(2.33)}   & 3.21  & \multicolumn{1}{l|}{(.74)}  & 1992.00         & (2554.83)         \\
	\multicolumn{1}{l|}{2}                                                 & 5447.24  & \multicolumn{1}{l|}{(2818.35)}  & 19.20  & \multicolumn{1}{l|}{(10.98)} & 229.19  & \multicolumn{1}{l|}{(23.98)} & 40.14   & \multicolumn{1}{l|}{(4.67)}   & 3.67  & \multicolumn{1}{l|}{(.59)}  & 2523.90         & (1451.65)         \\
	\multicolumn{1}{l|}{3}                                                 & 8511.04  & \multicolumn{1}{l|}{(4570.50)}  & 29.60  & \multicolumn{1}{l|}{(11.05)} & 218.54  & \multicolumn{1}{l|}{(73.25)} & 39.55   & \multicolumn{1}{l|}{(2.51)}   & 4.35  & \multicolumn{1}{l|}{(.80)}   & 5391.80         & (2603.08)         \\
	\multicolumn{1}{l|}{4}                                                 & 17872.71 & \multicolumn{1}{l|}{(7343.35)}  & 58.44  & \multicolumn{1}{l|}{(27.58)} & 263.12  & \multicolumn{1}{l|}{(59.19)} & 41.46   & \multicolumn{1}{l|}{(5.01)}   & 4.23  & \multicolumn{1}{l|}{(.75)}  & 10557.78        & (6438.14)         \\
	\multicolumn{1}{l|}{5}                                                 & 23059.11 & \multicolumn{1}{l|}{(4619.78)}  & 76.44  & \multicolumn{1}{l|}{(13.79)} & 237.34  & \multicolumn{1}{l|}{(71.62)} & 42.40   & \multicolumn{1}{l|}{(4.27)}   & 4.78  & \multicolumn{1}{l|}{(.89)} & 13470.00        & (2508.85)         \\
	\multicolumn{1}{l|}{6}                                                 & 24624.64 & \multicolumn{1}{l|}{(4802.55)}  & 84.10  & \multicolumn{1}{l|}{(43.68)} & 198.08  & \multicolumn{1}{l|}{(23.47)} & 41.80   & \multicolumn{1}{l|}{(3.46)}   & 5.16  & \multicolumn{1}{l|}{(1.10)} & 15714.80        & (8111.69)         \\ \hline \hline
	\multicolumn{1}{l|}{7}                                                 & 4714.34  & \multicolumn{1}{l|}{(1648.54)}  & 14.22  & \multicolumn{1}{l|}{(4.92)}  & 251.46  & \multicolumn{1}{l|}{(66.00)} & 35.73   & \multicolumn{1}{l|}{(5.26)}   & 2.75  & \multicolumn{1}{l|}{(.44)} & 897.33          & (222.74)          \\
	\multicolumn{1}{l|}{8}                                                 & 5805.69  & \multicolumn{1}{l|}{(1460.25)}  & 21.60  & \multicolumn{1}{l|}{(5.22)}  & 216.15  & \multicolumn{1}{l|}{(53.62)} & 39.24   & \multicolumn{1}{l|}{(4.25)}   & 3.67  & \multicolumn{1}{l|}{(.70)}  & 2491.50         & (1016.77)         \\
	\multicolumn{1}{l|}{9}                                                 & 7845.84  & \multicolumn{1}{l|}{(2530.96)}  & 29.10  & \multicolumn{1}{l|}{(8.58)}  & 201.69  & \multicolumn{1}{l|}{(41.57)} & 41.45   & \multicolumn{1}{l|}{(5.12)}   & 3.65  & \multicolumn{1}{l|}{(.33)}  & 3646.50         & (1242.66)         \\
	\multicolumn{1}{l|}{10}                                                & 18050.78 & \multicolumn{1}{l|}{(16979.42)} & 67.00  & \multicolumn{1}{l|}{(62.48)} & 210.52  & \multicolumn{1}{l|}{(34.52)} & 40.44   & \multicolumn{1}{l|}{(2.91)}   & 4.50  & \multicolumn{1}{l|}{(.68)}  & 10785.70        & (10120.79)        \\
	\multicolumn{1}{l|}{11}                                                & 25271.66 & \multicolumn{1}{l|}{(13222.79)} & 92.50  & \multicolumn{1}{l|}{(45.91)} & 234.99  & \multicolumn{1}{l|}{(39.24)} & 41.90   & \multicolumn{1}{l|}{(2.47)}   & 4.56  & \multicolumn{1}{l|}{(.52)}  & 14289.80        & (8291.09)         \\
	\multicolumn{1}{l|}{12}                                                & 37319.84 & \multicolumn{1}{l|}{(18296.72)} & 118.44 & \multicolumn{1}{l|}{(62.85)} & 235.38  & \multicolumn{1}{l|}{(53.97)} & 41.18   & \multicolumn{1}{l|}{(2.62)}   & 4.30  & \multicolumn{1}{l|}{(.97)} & 21383.11        & (11571.97)        \\ \hline \hline
	\multicolumn{1}{l|}{13}                                                & 4554.28  & \multicolumn{1}{l|}{(1763.95)}  & 14.90  & \multicolumn{1}{l|}{(6.74)}  & 270.74  & \multicolumn{1}{l|}{(65.02)} & 35.17   & \multicolumn{1}{l|}{(4.01)}   & 2.34  & \multicolumn{1}{l|}{(.72)}  & 1268.90         & (954.61)          \\
	\multicolumn{1}{l|}{14}                                                & 7152.53  & \multicolumn{1}{l|}{(2133.98)}  & 26.78  & \multicolumn{1}{l|}{(9.54)}  & 226.20  & \multicolumn{1}{l|}{(35.68)} & 40.10   & \multicolumn{1}{l|}{(5.35)}   & 2.86  & \multicolumn{1}{l|}{(.75)}  & 2713.22         & (990.86)          \\
	\multicolumn{1}{l|}{15}                                                & 11063.81 & \multicolumn{1}{l|}{(3868.80)}  & 40.78  & \multicolumn{1}{l|}{(14.32)} & 203.46  & \multicolumn{1}{l|}{(44.29)} & 41.50   & \multicolumn{1}{l|}{(3.90)}   & 3.12  & \multicolumn{1}{l|}{(.32)}  & 4953.22         & (1459.51)         \\
	\multicolumn{1}{l|}{16}                                                & 11712.67 & \multicolumn{1}{l|}{(4144.44)}  & 42.10  & \multicolumn{1}{l|}{(10.24)} & 177.47  & \multicolumn{1}{l|}{(37.79)} & 42.93   & \multicolumn{1}{l|}{(4.22)}   & 3.50  & \multicolumn{1}{l|}{(.45)}  & 5632.88         & (1985.05)         \\
	\multicolumn{1}{l|}{17}                                                & 12726.65 & \multicolumn{1}{l|}{(3498.27)}  & 50.20  & \multicolumn{1}{l|}{(13.99)} & 188.29  & \multicolumn{1}{l|}{(25.24)} & 43.30   & \multicolumn{1}{l|}{(7.25)}   & 4.71  & \multicolumn{1}{l|}{(.80)}  & 8395.60         & (2024.88)         \\
	\multicolumn{1}{l|}{18}                                                & 21975.60 & \multicolumn{1}{l|}{(8620.21)}  & 79.00  & \multicolumn{1}{l|}{(30.52)} & 214.33  & \multicolumn{1}{l|}{(35.85)} & 42.71   & \multicolumn{1}{l|}{(4.89)}   & 4.08  & \multicolumn{1}{l|}{(.95)}  & 12534.70        & (4563.27)         \\ \hline
	\multicolumn{13}{l}{Note: PM = Process model; Fix = Fixation; Sacc = Saccade; Amp = Amplitude}                                                                                                                                                                                                                            
\end{tabular}
	\label{eyedcsr}
\end{table*}

From the table, an increase in the duration as well as number of fixations and saccades is associated with an increase in process model complexity. Moreover, noticeable differences in the respective model mappings are recognizable (e.g., high duration for process model 12 and an overall higher fixation number in the upward mapping). In fixation duration, no discernible differences can be found, whereas in the number of saccades, which increased with complexity level, process model 12 shows a higher number. No differences are present in saccade duration, but, in turn, several differences in saccade amplitudes between model complexity as well as mapping are discernible. Finally, scan paths are increasing with process model complexity level, whereas the downward mapping represents the shortest paths in general. 

In Table~\ref{cqdscr}, for each process model (i.e., PM), the maximum achievable score in the comprehension questions (i.e., Max) is shown. Furthermore, mean and standard deviation for the scores (i.e., Correct; relative frequency) and response times (in ms) for answering respective comprehension questions as well as the entirety (i.e., All) thereof are presented for each process model mapping.

\begin{table}[]
	\centering
	\caption{Descriptive results regarding the comprehension questions and response times}
\begin{tabular}{l|l|l|l|ll|l}
	PM & Max & \multicolumn{2}{c|}{Correct} & \multicolumn{2}{c|}{Response Time (ms)} & All       \\ \hline \hline
	1  & 10  & 1.00         & (.00)         & 3487.00            & (1401.11)          &           \\
	2  & 10  & .90         & (.00)         & 6515.00            & (3669.91)          &           \\
	3  & 20  & 1.60         & (.66)         & 7611.50            & (5067.18)          &           \\
	4  & 18  & 1.56         & (.68)         & 11072.22           & (5697.64)          &           \\
	5  & 27  & 2.11         & (.74)         & 7579.89            & (5418.33)          & 7001.40   \\
	6  & 30  & 2.10         & (.94)         & 5754.81            & (4507.46)          & (2289.20) \\ \hline \hline
	7  & 9   & 1.00         & (.00)         & 4144.44            & (2061.05)          &           \\
	8  & 10  & .90         & (.30)         & 5872.00            & (2677.13)          &           \\
	9  & 20  & 1.00         & (.41)         & 4803.00            & (2689.96)          &           \\
	10 & 20  & .80          & (.78)         & 7529.55            & (5026.06)          &           \\
	11 & 30  & 1.20         & (.81)         & 10478.15           & (5713.86)          & 6987.61   \\
	12 & 27  & 1.31         & (.84)         & 9098.52            & (5755.81)          & (2274.28) \\ \hline \hline
	13 & 10  & .90         & (.30)         & 4357.00            & (1368.07)          &           \\
	14 & 9   & .90         & (.30)         & 9013.33            & (3992.20)          &           \\
	15 & 18  & 1.77         & (.47)         & 5268.33            & (2443.72)          &           \\
	16 & 20  & 1.60         & (.90)         & 6448.00            & (3457.46)          &           \\
	17 & 30  & 2.30         & (.82)         & 8185.81            & (3756.28)          & 6812.16   \\ 
	18 & 30  & 2.50         & (.71)         & 7601.48            & (4677.45)          & (1627.69) \\ \hline \hline
	\multicolumn{7}{l}{\textbf{Note:} PM = Process model}  
\end{tabular}
\label{cqdscr}
\end{table}

Regarding the comprehension question scores, it is recognizable that in process models reflecting an easy level of complexity almost all participants answered correctly. However, with increasing complexity level, the number of correct answers decreases. Of particular interest are the results regrading the upward mapping. Whereas the results for the straight and downward mapping are similar, they differ in the upward mapping (i.e., fewer correct answers with increasing process model complexity). Regarding the response times for answering, no major differences are discernible and participants needed the same amount of time for answering the comprehension questions in respective mapping type. However, concerning model complexity, it appears that participants needed more time for answering the questions with increasing level of process model complexity.

Finally, results (i.e., mean and standard deviation) representing the three cognitive load dimensions (i.e., intrinsic, extraneous, germane) and, regarding the level of acceptability, perceived usefulness for understandability (i.e., PUU) as well as perceived ease of understandability (i.e., PEU) are shown in Table~\ref{cltdscr}. Further, the table includes the absolute frequency of the indicated preference regarding process model mapping.

\begin{table}[]
	\centering
	\caption{Descriptive results regarding the cognitive load, level of acceptability, and process model mapping preference}
	\label{cltdscr}
	\begin{tabular}{ccc}
		\multicolumn{3}{c}{Cognitive Load}                                                                   \\ \hline \hline
		\multicolumn{1}{c|}{Intrinsic}          & \multicolumn{1}{c|}{Extraneous}         & \multicolumn{1}{c}{Germane}       \\ \hline
		\multicolumn{1}{l|}{3.52 (1.62)}  & \multicolumn{1}{l|}{3.62 (1.55)} & 3.38 (1.64)                   \\ \hline \hline
		\multicolumn{3}{c}{Level of Acceptability}                                                           \\ \hline
		\multicolumn{1}{c|}{PUU}          & \multicolumn{1}{c|}{}            & \multicolumn{1}{c}{PEU}       \\ \cline{1-1} \cline{3-3} 
		\multicolumn{1}{l|}{19.76 (5.05)} & \multicolumn{1}{l|}{}            & 21.07 (5.21)                   \\ \hline \hline
		\multicolumn{3}{c}{Mapping Preference}                                                               \\ \hline
		\multicolumn{1}{c|}{Straight}     & \multicolumn{1}{c|}{Upward}     & \multicolumn{1}{c}{Downward} \\ \hline
		\multicolumn{1}{c|}{18}           & \multicolumn{1}{c|}{3}           & 8                             \\ \hline \hline
	\end{tabular}
\end{table}

\begin{table*}[!h]
	\centering
	\scriptsize
	\caption{Results of inferential statistics for considered performance measures}
	\label{inf}
	\begin{tabular}{llllllllllll}
		\cline{2-4} \cline{6-8} \cline{10-12}
		\multicolumn{1}{l|}{} & \multicolumn{3}{c|}{Comprehension Duration}                                                     & \multicolumn{1}{l|}{} & \multicolumn{3}{c|}{Number of Fixations}                                              & \multicolumn{1}{l|}{} & \multicolumn{3}{c|}{Average Fixation Duration}                           \\ \cline{1-4} \cline{6-8} \cline{10-12} 
		\multicolumn{1}{|l|}{ME 1}                                           & F(5; 45)              & = 31.55;                  & \multicolumn{1}{l|}{p = .001} & \multicolumn{1}{l|}{} & F(5; 45)                      & = 36.78;              & \multicolumn{1}{l|}{p = .001} & \multicolumn{1}{l|}{} & F(2.52; 22.66)                & = 6.32;  & \multicolumn{1}{l|}{p = .004} \\ \cline{1-4} \cline{6-8} \cline{10-12} 
		\multicolumn{1}{|l|}{ME 2}                                           & F(1.41; 12.67)        & = 8.31;                   & \multicolumn{1}{l|}{p = .008} & \multicolumn{1}{l|}{} & F(1.57; 14.11)                & = 8.20;               & \multicolumn{1}{l|}{p = .006} & \multicolumn{1}{l|}{} & F(2; 18)                      & = 2.11;  & \multicolumn{1}{l|}{p = .150} \\ \cline{1-4} \cline{6-8} \cline{10-12} 
		\multicolumn{1}{|l|}{IE}                                             & F(2.91; 26.20)        & = 2.57;                   & \multicolumn{1}{l|}{p = .078} & \multicolumn{1}{l|}{} & F(2.68; 24.09)                & = 3.38;               & \multicolumn{1}{l|}{p = .039} & \multicolumn{1}{l|}{} & F(3.30; 29.69)                & = 1.94;  & \multicolumn{1}{l|}{p = .140} \\ \cline{1-4} \cline{6-8} \cline{10-12} 
		&                       &                           &                               &                       &                               &                       &                               &                       &                               &          &                               \\ \cline{2-4} \cline{6-8} \cline{10-12} 
		\multicolumn{1}{l|}{}                                                & \multicolumn{3}{c|}{Average Saccade Duration}                                     & \multicolumn{1}{l|}{} & \multicolumn{3}{c|}{Average Saccade Amplitude}                                        & \multicolumn{1}{l|}{} & \multicolumn{3}{c|}{Scan Path}                                           \\ \cline{1-4} \cline{6-8} \cline{10-12} 
		\multicolumn{1}{|l|}{ME 1}                                           & F(5; 45)              & = 9.19;                   & \multicolumn{1}{l|}{p = .001} & \multicolumn{1}{l|}{} & F(5; 45)                      & = 28.13;              & \multicolumn{1}{l|}{p = .001} & \multicolumn{1}{l|}{} & F(1.63; 14.63)                & = 37.56; & \multicolumn{1}{l|}{p = .001} \\ \cline{1-4} \cline{6-8} \cline{10-12} 
		\multicolumn{1}{|l|}{ME 2}                                           & F(2; 18)              & = .89;                    & \multicolumn{1}{l|}{p = .427} & \multicolumn{1}{l|}{} & F(2; 18)                      & = 13.10;              & \multicolumn{1}{l|}{p = .001} & \multicolumn{1}{l|}{} & F(2; 18)                      & = 7.43;  & \multicolumn{1}{l|}{p = .004} \\ \cline{1-4} \cline{6-8} \cline{10-12} 
		\multicolumn{1}{|l|}{IE}                                             & F(10; 90)             & = .26;                    & \multicolumn{1}{l|}{p = .988} & \multicolumn{1}{l|}{} & F(10; 90)                     & = .94;                & \multicolumn{1}{l|}{p = .500} & \multicolumn{1}{l|}{} & F(1.73; 15.57)                & = 2.09;  & \multicolumn{1}{l|}{p = .161} \\ \cline{1-4} \cline{6-8} \cline{10-12} 
		&                       &                           &                               &                       &                               &                       &                               &                       &                               &          &                               \\ \cline{4-6} \cline{8-10}
		& \multicolumn{1}{c}{}  & \multicolumn{1}{l|}{}     & \multicolumn{3}{c|}{Comprehension Score}                                                            & \multicolumn{1}{c|}{} & \multicolumn{3}{c|}{Response Time}                                                    &          &                               \\ \cline{3-6} \cline{8-10}
		& \multicolumn{1}{l|}{} & \multicolumn{1}{l|}{ME 1} & F(5; 45)                      & = 18.20;              & \multicolumn{1}{l|}{p = .001} & \multicolumn{1}{l|}{} & F(1.62; 14.57)                & = 10.86;              & \multicolumn{1}{l|}{p = .002} &          &                               \\ \cline{3-6} \cline{8-10}
		& \multicolumn{1}{l|}{} & \multicolumn{1}{l|}{ME 2} & F(2; 18)                      & = 13.85;              & \multicolumn{1}{l|}{p = .001} & \multicolumn{1}{l|}{} & F(2; 18)                      & = .06;                & \multicolumn{1}{l|}{p = .947} &          &                               \\ \cline{3-6} \cline{8-10}
		& \multicolumn{1}{l|}{} & \multicolumn{1}{l|}{IE}   & F(10; 90)                     & = 2.39;               & \multicolumn{1}{l|}{p = .015} & \multicolumn{1}{l|}{} & F(3.54; 31.90)                & = 3.13;               & \multicolumn{1}{l|}{p = .032} &          &                               \\ \cline{3-6} \cline{8-10}
		\multicolumn{12}{l}{\textbf{Note:} ME = Main effect, IE = Interaction effect}
	\end{tabular}
\end{table*}

In general, the findings regarding the cognitive load dimensions represent a moderate level. More specifically, despite the plain structure of the process models (i.e., sequence), the inherent complexity of the presented model mappings was challenging to cope with for the participants. Further, participants needed an average amount of expend for the construction of mental models during comprehension. The results concerning PUU and PEU (i.e., above average for both) indicate that, on the one hand, the application of process models in practice (e.g., for the communication of information) may be beneficial. On the other hand, there is a positive attitude supposing that the comprehension of the presented process model mappings is associated with less mental effort. Finally, regarding mapping preference, the straight model mapping is the most congenial mapping in terms of comprehension based on subjective preferences, whereas the upward mapping received the least consent. Considering the results in the Tables~\ref{eyedcsr} and \ref{cqdscr}, although participants stated that the straight mapping was the best comprehensible mapping type, however, attention was conveyed more effectively in the downward mapping (e.g., fewer fixations with higher comprehension score).  

\subsection{Inferential Statistics}
\label{inferential}

Results presented in Section~\ref{descriptive} are merely based on descriptive differences. In order to evaluate whether the differences reported in the descriptive results reached statistical significance, analyses of variances (ANOVAs) were performed for all performance measures (see Section~\ref{measures}). Thereby, missing values in each performance measure for the ANOVA (i.e., 6 from 180 (3,33 \%)) were imputed using Expectation-Maximization (i.e., non-significant Little's MCAR test: duration (p = .932), number of fixations (p = .799), average fixation duration (p = .996), number of saccades (p = .902), average saccade duration (p = .892), average saccade amplitude (p = .977), scan path (p = .946), score (p = .775), response time (p = .780)). Moreover, Greenhouse-Geisser correction was applied if necessary (i.e., significant Mauchly's sphericity test). Thereby, two within-subject factors "process model complexity" (six levels: process model 1 - 6) and "mapping" (three levels: straight (i.e., 1), upwards (i.e., 2), downwards (i.e., 3)) were examined. The main effects for the process model complexity 1 - 6 (i.e., ME 1) and for the mapping 1 - 3  (i.e., ME 2) were evaluated as well as the interaction effect complexity*mapping (i.e., IE). In addition, in the event of significance for ME 1, repeated contrasts were employed. In the event of significance for ME 2, t-tests for paired samples between each level of "complexity" and "mapping" were performed (i.e., bonferroni correction). To better explain the IE in the event of significance, pairwise comparisons of estimated marginal means (i.e., bonferroni correction) between all six levels (i.e., process model 1 - 6) of "complexity" for each "mapping" were performed. Finally, all statistical tests were performed two-tailed and the significance value was set to p \textless{} .05. Table~\ref{inf} presents the results obtained of inferential statistics for all performance measures. 

Regarding \textbf{comprehension duration}, ME 1 was significant and repeated contrasts showed that complexity 2 (p = .062) did not have a longer duration (m = 6104.94 (2325.70)) than complexity 1 (m = 4844.31 (2363.99)), but complexity 3 (p = .007) had a longer duration (m = 9114.32 (3992.55)) than complexity 2. Complexity 4 (p = .002) had a longer duration (m = 15804.25 (11441.77)) than complexity 3, whereas complexity 5 (p = .054) did not have a longer duration (m = 19925.26 (9928.16)) than complexity 4. Complexity 6 (p = .007) had a longer duration (m = 27876.01 (15807.30)) than complexity 5. 
ME 2 reached significance and the t-tests for paired samples reached significance in complexity 5 between mapping 1 and 3 (p = .001; longer durations in mapping 1) as well as mapping 2 and 3 (p = .015; longer durations in mapping 2).
IE reached no significance.

Regarding \textbf{number of fixations}, ME 1 was significant and repeated contrasts showed that complexity 2 (p = .002) had more fixations (m = 22.40 (9.43)) than complexity 1 (m = 14.22 (6.02)) and complexity 3 (p = .003) had more fixations (m = 32.84 (12.62)) than complexity 2. Complexity 4 (p = .010) had more fixations (m = 55.31 (41.65)) than complexity 3 and complexity 5 (p = .003) had more fixations (m = 72.55 (38.63)) than complexity 4. Complexity 6 (p = .003) had more fixations (m = 101.05 (53.14)) than complexity 5. 
ME 2 reached significance and the t-tests for paired samples reached significance in complexity 5 between mapping 1 and 3 (p = .004; more fixations in mapping 1). In complexity 6 between mapping 1 and 2 (p = .014; more fixations in mapping 2) as well as mapping 2 and 3 (p = .009; more fixations in mapping 2).
IE reached significance and the pairwise comparisons of estimated marginal means indicated a significance in complexity 5 between mapping 1 and 3 (p = .012). In complexity 6 between mapping 1 and 2 (p = .043) as well as mapping 2 and 3 (p = .026).

Regarding \textbf{average fixation duration}, ME 1 was significant and repeated contrasts showed that complexity 2 (p = .030) had a shorter fixation duration (m = 224.03 (40.24)) than complexity 1 (m = 257.27 (65.33)), whereas complexity 3 (p = .090) did not have a shorter fixation duration (m = 208.80 (55.95)) than complexity 2. Complexity 4 (p = .535) did not have a shorter fixation duration (m = 214.92 (56.66)) than complexity 3 and complexity 5 (p = .489) did not have a shorter fixation duration (m = 219.57 (53.48)) than complexity 4. Complexity 6 (p = .393) did not have a shorter fixation duration (m = 214.89 (42.06)) than complexity 5. Further, neither ME 2 nor IE reached significance.

Regarding \textbf{average saccade duration}, ME 1 was significant and repeated contrasts showed that complexity 2 (p = .012) had a longer saccade duration (m = 39.77 (6.21)) than complexity 1 (m = 35.60 (4.61)), whereas complexity 3 (p = .514) did not have a longer saccade duration (m = 40.85 (4.10)) than complexity 2. Complexity 4 (p = .545) did not have a longer saccade duration (m = 41.54 (4.26)) than complexity 3 and complexity 5 (p = .441) did not have a longer saccade duration (m = 42.47 (5.14)) than complexity 4. Complexity 6 (p = .656) did not have a longer saccade duration (m = 41.92 (3.86)) than complexity 5. Further, neither ME 2 nor IE reached significance.

Regarding \textbf{average saccade amplitude}, ME 1 was significant and repeated contrasts showed that complexity 2 (p = .016) had longer saccade amplitudes (m = 3.41 (.78)) than complexity 1 (m = 2.77 (1.03)) and complexity 3 (p = .046) had longer saccade amplitudes (m = 3.72 (.73)) than complexity 2. Complexity 4 (p = .017) had longer saccade amplitudes (m = 4.09 (.77)) than complexity 3 and complexity 5 (p = .007) had longer saccade amplitudes (m = 5.16 (2.78)) than complexity 4. Complexity 6 (p = .604) did not have longer saccade amplitudes (m = 4.51 (1.02)) than complexity 5.
ME 2 reached significance and the t-tests for paired samples reached significance in complexity 3 between mapping 1 and 3 (p = .002; longer amplitudes in mapping 1) as well as mapping 2 and 3 (p = .003; longer amplitudes in mapping 2). In complexity 4 between mapping 2 and 3 (p = . 005; longer amplitudes in mapping 2).
IE reached no significance.

Regarding \textbf{scan path}, ME 1 was significant and repeated contrasts showed that complexity 2 (p = .007) had a longer scan path (m = 2575.33 (1182.16)) than complexity 1 (m = 1372.83 (1677.12)) and complexity 3 (p = .001) had a longer scan path (m = 4657.96 (2023.89)) than complexity 2. Complexity 4 (p = .007) had a longer scan path (m = 8963.04 (7405.62)) than complexity 3 and complexity 5 (p = .009) did had a longer scan path (m = 11928.07 (5847.90)) than complexity 4. Complexity 6 (p = .008) had a longer scan path (m = 16573.96 (9252.44)) than complexity 5. 
ME 2 reached significance and the t-tests for paired samples reached significance in complexity 5 between mapping 1 and 3 (p = .001; longer scan paths in mapping 1).
IE reached no significance.

Regarding \textbf{comprehension score}, ME 1 was significant and repeated contrasts showed that complexity 2 (p = .443) did not have a higher score (m = .90 (.31)) than complexity 1 (m = .97 (.18)), but complexity 3 (p = .004) had a higher score (m = 1.46 (.73)) than complexity 2. Complexity 4 (p = .541) did not have a higher score (m = 1.35 (.76)) than complexity 3, but complexity 5 (p = .011) had a higher score (m = 1.90 (.92)) than complexity 4. Complexity 6 (p = .743) did not have a higher score (m = 1.97 (.96) than complexity 5.
ME 2 reached significance and the t-tests for paired samples reached significance in complexity 5 between mapping 2 and 3 (p = . 012; lower scores in mapping 2). In complexity 6 between mapping 2 and 3 (p = .016; lower scores in mapping 2).
IE reached significance and the pairwise comparisons of estimated marginal means indicated a significance in complexity 5 between mapping 2 and 3 (p = .035). In complexity 6 between mapping 2 and 3 (p = .050).

Regarding \textbf{response time}, ME 1 was significant and repeated contrasts showed that complexity 2 (p = .001) had a longer response time (m = 7096.49 (3722.67)) than complexity 1 (m = 4005.59 (1669.76)), whereas complexity 3 (p = .119) did not have a longer response time (m = 5893.71 (2600.42)) than complexity 2. Complexity 4 (p = .008) had a longer response time (m = 8279.32 (4431.67)) than complexity 3, whereas complexity 5 (p = .362) did not have a longer response time (m = 8609.79 (3512.15)) than complexity 4. Complexity 6 (p = .002) had a shorter response time (m = 7267.24 (3727.66)) than complexity 5.
ME 2 reached no significance. 
IE reached significance but the pairwise comparisons of estimated marginal means indicated no significant differences (e.g., due the lack of statistical power).

\subsection{Further Eye Movement Analyses}
\label{eye}

Based on inferential statistics (see Section~\ref{inferential}), several significant differences are identified in the performed eye movements from the participants during the comprehension of the process models in the study. In order to get a better understanding of attention shifts (i.e., information extraction) in the process models by means of fixations, regions of interest (i.e., ROIs) are defined for each process model element (i.e., events and activities) across all models. Thereby, a ROI constitutes a manually defined area in a stimulus, which can be used for the calculation of specific ROI-based metrics (e.g., ROI hit, entry, time, and revisits) \cite{carter2020best}. The following Tables~\ref{crazy1}, \ref{crazy2}, and \ref{crazy3} present for each element (i.e., S = start event; E = end event; activities numerically ascending from left to right) in respective process models from the study (i.e., based on level of complexity and mapping), the ROI-based metrics number of fixations (i.e., absolute frequency) and the number of fixation hits. Regarding the latter, it shows whether a participants' attention (i.e., fixation) was placed on this model element. Moreover, concerning the process models with an upward or downward mapping, the sequences, which were displayed horizontally (i.e., H) and vertically (i.e., V), were defined as two separate ROIs.

Regarding the \textbf{easy process models} (see Table~\ref{crazy1}), similarities as well as differences in the number of fixations can be seen. More specifically, while all activities from the process models have been considered, the two events (i.e., start and end) were not considered by all participants. In this regard, the events did not reflect any semantic information (i.e., empty labels), but syntactically marked the beginning and the end of a process model. However, some comprehension questions were concerned with the first and last activity in a model (e.g., activity M is the first/last activity in the process model) and, thus, the events may constitute reference points for the identification of respective activities. Since not all participants fixated the events but still answered respective comprehension questions correctly, the identification had to be done differently by these participants. An explanation may be that the identification of the start and end event was made directly at the first sight of the process model through their planar structure. More specifically, due to the size of the models and the distance to the monitor, participants were able to observe the process models as a whole (i.e., within the field of view at different levels of peripheral vision \cite{simpson2017mini}) and, hence, the planar structures (i.e., circle vs. rectangle) were separated at first glance. Thus, it was no longer necessary to consider the events and, as a result, the participants were able to allocate the cognitive resources for comprehension on other model elements. In addition, a confirmation of a correct separation between activities and events was obtained through the peripheral vision (i.e., indirect fixation). Finally, the two ROIs, separating the horizontal and vertical sequences in the upward as well as downward mapping, were equally fixated. 

Regarding the \textbf{medium process models} (see Table~\ref{crazy2}), similar as in the easy models (see Table~\ref{crazy1}), the start and the end event were not fixated by all participants. In turn, these two events had been considered even less, demonstrating that not relevant elements in a process models are not considered with increasing level of model complexity. Moreover, in the straight and downward mapping, it can be seen that the number of fixations on the individual activities increases at the beginning and then decreases so that the posterior activities were fixated less frequently. An explanation for this observation may be that with the start of the fixations on the activities, a saturation effect occurred during process model comprehension. More specifically, the working memory may be confronted with capacity limitations during the comprehension of the information from the first activities (i.e., magical number seven \cite{corbett2017magical}). As a result, information from the latter activities were not fixated to the same extent, as related capacities were exhausted. Furthermore, participants may recognize that planar information (e.g., structure) in the process models remained and did not change. As a consequence, attention shifted only between activity labels. While the vertical ROI was fixated less frequently in the downward mapping (i.e., due to the saturation effect), the number of fixations in the two ROIs (i.e., horizontal and vertical) was approximately equal in the upward mapping. The saturation effect, as seen in the straight and downward mapping, was not present in this mapping type. 

Regarding the \textbf{hard process models} (see Table~\ref{crazy3}), the same observations as in the both prior levels of complexity (i.e., easy (see Table~\ref{crazy1}) and medium (see Table~\ref{crazy2})) can be seen. In more detail, events were not fixated by all participants and a saturation effect was noticeable in the process models reflecting a straight and downward mapping, but not in the upward mapping. Considering the score (see Table~\ref{cqdscr}), the findings confirm that the visual system was conveyed more effectively during the comprehension of process models with a downward mapping (i.e., higher average score).   

\begin{table}[]
	\caption{ROIs in process models with easy complexity level}
	\centering
	\scriptsize
\begin{tabular}{ccccccccccc}
	\multicolumn{3}{c}{Straight} &  & \multicolumn{3}{c}{Upward}                                   &  & \multicolumn{3}{c}{Downward}                                 \\ \cline{1-3} \cline{5-7} \cline{9-11} 
	\multicolumn{3}{c}{Process Model 1}                                         &  & \multicolumn{3}{c}{Process Model 7}                          &  & \multicolumn{3}{c}{Process Model 13}                         \\ \cline{1-3} \cline{5-7} \cline{9-11} 
	\multicolumn{1}{l|}{ROI}      & \multicolumn{1}{l|}{No Fix}      & Hit      &  & \multicolumn{1}{l|}{ROI} & \multicolumn{1}{l|}{No Fix} & Hit &  & \multicolumn{1}{l|}{ROI} & \multicolumn{1}{l|}{No Fix} & Hit \\ \cline{1-3} \cline{5-7} \cline{9-11} 
	\multicolumn{1}{l|}{S}        & \multicolumn{1}{l|}{11}          & 5        &  & \multicolumn{1}{l|}{S}   & \multicolumn{1}{l|}{14}     & 7   &  & \multicolumn{1}{l|}{S}   & \multicolumn{1}{l|}{26}     & 8   \\
	\multicolumn{1}{l|}{1}        & \multicolumn{1}{l|}{79}          & 10       &  & \multicolumn{1}{l|}{1}   & \multicolumn{1}{l|}{75}     & 9   &  & \multicolumn{1}{l|}{1}   & \multicolumn{1}{l|}{69}     & 10  \\
	\multicolumn{1}{l|}{E}        & \multicolumn{1}{l|}{22}          & 9        &  & \multicolumn{1}{l|}{E}   & \multicolumn{1}{l|}{33}     & 9   &  & \multicolumn{1}{l|}{E}   & \multicolumn{1}{l|}{30}     & 7   \\
	\multicolumn{1}{l|}{-}        & \multicolumn{1}{l|}{-}           & -        &  & \multicolumn{1}{l|}{H}   & \multicolumn{1}{l|}{96}     & 9   &  & \multicolumn{1}{l|}{H}   & \multicolumn{1}{l|}{105}    & 10  \\
	\multicolumn{1}{l|}{-}        & \multicolumn{1}{l|}{-}           & -        &  & \multicolumn{1}{l|}{V}   & \multicolumn{1}{l|}{106}    & 9   &  & \multicolumn{1}{l|}{V}   & \multicolumn{1}{l|}{101}    & 9   \\
	&                                  &          &  &                          &                             &     &  &                          &                             &     \\
	\multicolumn{3}{c}{Process Model 2}                                         &  & \multicolumn{3}{c}{Process Model 8}                          &  & \multicolumn{3}{c}{Process Model 14}                         \\ \cline{1-3} \cline{5-7} \cline{9-11} 
	\multicolumn{1}{l|}{ROI}      & \multicolumn{1}{l|}{No Fix}      & Hit      &  & \multicolumn{1}{l|}{ROI} & \multicolumn{1}{l|}{No Fix} & Hit &  & \multicolumn{1}{l|}{ROI} & \multicolumn{1}{l|}{No Fix} & Hit \\ \cline{1-3} \cline{5-7} \cline{9-11} 
	\multicolumn{1}{l|}{S}        & \multicolumn{1}{l|}{7}           & 5        &  & \multicolumn{1}{l|}{S}   & \multicolumn{1}{l|}{9}      & 4   &  & \multicolumn{1}{l|}{S}   & \multicolumn{1}{l|}{6}      & 4   \\
	\multicolumn{1}{l|}{1}        & \multicolumn{1}{l|}{49}          & 10       &  & \multicolumn{1}{l|}{1}   & \multicolumn{1}{l|}{50}     & 10  &  & \multicolumn{1}{l|}{1}   & \multicolumn{1}{l|}{78}     & 9   \\
	\multicolumn{1}{l|}{2}        & \multicolumn{1}{l|}{68}          & 10       &  & \multicolumn{1}{l|}{2}   & \multicolumn{1}{l|}{43}     & 10  &  & \multicolumn{1}{l|}{2}   & \multicolumn{1}{l|}{62}     & 9   \\
	\multicolumn{1}{l|}{3}        & \multicolumn{1}{l|}{47}          & 10       &  & \multicolumn{1}{l|}{3}   & \multicolumn{1}{l|}{67}     & 10  &  & \multicolumn{1}{l|}{3}   & \multicolumn{1}{l|}{54}     & 9   \\
	\multicolumn{1}{l|}{E}        & \multicolumn{1}{l|}{9}           & 8        &  & \multicolumn{1}{l|}{E}   & \multicolumn{1}{l|}{10}     & 3   &  & \multicolumn{1}{l|}{E}   & \multicolumn{1}{l|}{12}     & 6   \\
	\multicolumn{1}{l|}{-}        & \multicolumn{1}{l|}{-}           & -        &  & \multicolumn{1}{l|}{H}   & \multicolumn{1}{l|}{117}    & 10  &  & \multicolumn{1}{l|}{H}   & \multicolumn{1}{l|}{154}    & 9   \\
	\multicolumn{1}{l|}{-}        & \multicolumn{1}{l|}{-}           & -        &  & \multicolumn{1}{l|}{V}   & \multicolumn{1}{l|}{126}    & 10  &  & \multicolumn{1}{l|}{V}   & \multicolumn{1}{l|}{136}    & 9  \\
	\multicolumn{11}{l}{\textbf{Note:} ROI = Region of interest; Fix = Fixation; activities are numerically ascending}  \\
\multicolumn{11}{l}{from left to right; S = Start event; E = End event; H = Horizontal; V = Vertical}  
\end{tabular}
	\label{crazy1}
\end{table}

\begin{table}[]
	\caption{ROIs in process models with medium complexity level}
	\centering
	\scriptsize
	\begin{tabular}{ccccccccccc}
		\multicolumn{3}{c}{Straight} &  & \multicolumn{3}{c}{Upward}                                   &  & \multicolumn{3}{c}{Downward}                                 \\ \cline{1-3} \cline{5-7} \cline{9-11} 
		\multicolumn{3}{c}{Process Model 3}                                         &  & \multicolumn{3}{c}{Process Model 9}                          &  & \multicolumn{3}{c}{Process Model 15}                         \\ \cline{1-3} \cline{5-7} \cline{9-11} 
		\multicolumn{1}{c|}{ROI}      & \multicolumn{1}{c|}{No Fix}      & Hit      &  & \multicolumn{1}{c|}{ROI} & \multicolumn{1}{c|}{No Fix} & Hit &  & \multicolumn{1}{c|}{ROI} & \multicolumn{1}{c|}{No Fix} & Hit \\ \cline{1-3} \cline{5-7} \cline{9-11} 
		\multicolumn{1}{c|}{S}        & \multicolumn{1}{c|}{3}           & 1        &  & \multicolumn{1}{c|}{S}   & \multicolumn{1}{c|}{5}      & 3   &  & \multicolumn{1}{c|}{S}   & \multicolumn{1}{c|}{4}      & 3   \\
		\multicolumn{1}{c|}{1}        & \multicolumn{1}{c|}{52}          & 10       &  & \multicolumn{1}{c|}{1}   & \multicolumn{1}{c|}{42}     & 10  &  & \multicolumn{1}{c|}{1}   & \multicolumn{1}{c|}{60}     & 9   \\
		\multicolumn{1}{c|}{2}        & \multicolumn{1}{c|}{83}          & 10       &  & \multicolumn{1}{c|}{2}   & \multicolumn{1}{c|}{60}     & 10  &  & \multicolumn{1}{c|}{2}   & \multicolumn{1}{c|}{99}     & 9   \\
		\multicolumn{1}{c|}{3}        & \multicolumn{1}{c|}{59}          & 9        &  & \multicolumn{1}{c|}{3}   & \multicolumn{1}{c|}{40}     & 9   &  & \multicolumn{1}{c|}{3}   & \multicolumn{1}{c|}{69}     & 8   \\
		\multicolumn{1}{c|}{4}        & \multicolumn{1}{c|}{42}          & 10       &  & \multicolumn{1}{c|}{4}   & \multicolumn{1}{c|}{55}     & 10  &  & \multicolumn{1}{c|}{4}   & \multicolumn{1}{c|}{43}     & 8   \\
		\multicolumn{1}{c|}{5}        & \multicolumn{1}{c|}{22}          & 10       &  & \multicolumn{1}{c|}{5}   & \multicolumn{1}{c|}{51}     & 10  &  & \multicolumn{1}{c|}{5}   & \multicolumn{1}{c|}{30}     & 9   \\
		\multicolumn{1}{c|}{E}        & \multicolumn{1}{c|}{1}           & 1        &  & \multicolumn{1}{c|}{E}   & \multicolumn{1}{c|}{1}      & 1   &  & \multicolumn{1}{c|}{E}   & \multicolumn{1}{c|}{5}      & 3   \\
		\multicolumn{1}{c|}{-}        & \multicolumn{1}{c|}{-}           & -        &  & \multicolumn{1}{c|}{H}   & \multicolumn{1}{c|}{158}    & 10  &  & \multicolumn{1}{c|}{H}   & \multicolumn{1}{c|}{245}    & 9   \\
		\multicolumn{1}{c|}{-}        & \multicolumn{1}{c|}{-}           & -        &  & \multicolumn{1}{c|}{V}   & \multicolumn{1}{c|}{154}    & 10  &  & \multicolumn{1}{c|}{V}   & \multicolumn{1}{c|}{161}    & 9   \\
		&                                  &          &  &                          &                             &     &  &                          &                             &     \\
		\multicolumn{3}{c}{Process Model 4}                                         &  & \multicolumn{3}{c}{Process Model 10}                         &  & \multicolumn{3}{c}{Process Model 16}                         \\ \cline{1-3} \cline{5-7} \cline{9-11} 
		\multicolumn{1}{c|}{ROI}      & \multicolumn{1}{c|}{No Fix}      & Hit      &  & \multicolumn{1}{c|}{ROI} & \multicolumn{1}{c|}{No Fix} & Hit &  & \multicolumn{1}{c|}{ROI} & \multicolumn{1}{c|}{No Fix} & Hit \\ \cline{1-3} \cline{5-7} \cline{9-11} 
		\multicolumn{1}{c|}{S}        & \multicolumn{1}{c|}{16}          & 1        &  & \multicolumn{1}{c|}{S}   & \multicolumn{1}{c|}{20}     & 4   &  & \multicolumn{1}{c|}{S}   & \multicolumn{1}{c|}{3}      & 2   \\
		\multicolumn{1}{c|}{1}        & \multicolumn{1}{c|}{74}          & 9        &  & \multicolumn{1}{c|}{1}   & \multicolumn{1}{c|}{85}     & 10  &  & \multicolumn{1}{c|}{1}   & \multicolumn{1}{c|}{52}     & 10  \\
		\multicolumn{1}{c|}{2}        & \multicolumn{1}{c|}{79}          & 9        &  & \multicolumn{1}{c|}{2}   & \multicolumn{1}{c|}{97}     & 10  &  & \multicolumn{1}{c|}{2}   & \multicolumn{1}{c|}{82}     & 10  \\
		\multicolumn{1}{c|}{3}        & \multicolumn{1}{c|}{58}          & 9        &  & \multicolumn{1}{c|}{3}   & \multicolumn{1}{c|}{83}     & 10  &  & \multicolumn{1}{c|}{3}   & \multicolumn{1}{c|}{57}     & 9   \\
		\multicolumn{1}{c|}{4}        & \multicolumn{1}{c|}{56}          & 9        &  & \multicolumn{1}{c|}{4}   & \multicolumn{1}{c|}{57}     & 10  &  & \multicolumn{1}{c|}{4}   & \multicolumn{1}{c|}{41}     & 9   \\
		\multicolumn{1}{c|}{5}        & \multicolumn{1}{c|}{50}          & 9        &  & \multicolumn{1}{c|}{5}   & \multicolumn{1}{c|}{83}     & 10  &  & \multicolumn{1}{c|}{5}   & \multicolumn{1}{c|}{42}     & 10  \\
		\multicolumn{1}{c|}{6}        & \multicolumn{1}{c|}{28}          & 7        &  & \multicolumn{1}{c|}{6}   & \multicolumn{1}{c|}{92}     & 10  &  & \multicolumn{1}{c|}{6}   & \multicolumn{1}{c|}{58}     & 10  \\
		\multicolumn{1}{c|}{7}        & \multicolumn{1}{c|}{31}          & 9        &  & \multicolumn{1}{c|}{7}   & \multicolumn{1}{c|}{87}     & 10  &  & \multicolumn{1}{c|}{7}   & \multicolumn{1}{c|}{35}     & 9   \\
		\multicolumn{1}{c|}{E}        & \multicolumn{1}{c|}{4}           & 3        &  & \multicolumn{1}{c|}{E}   & \multicolumn{1}{c|}{11}     & 5   &  & \multicolumn{1}{c|}{E}   & \multicolumn{1}{c|}{1}      & 1   \\
		\multicolumn{1}{c|}{-}        & \multicolumn{1}{c|}{-}           & -        &  & \multicolumn{1}{c|}{H}   & \multicolumn{1}{c|}{361}    & 10  &  & \multicolumn{1}{c|}{H}   & \multicolumn{1}{c|}{246}    & 10  \\
		\multicolumn{1}{c|}{-}        & \multicolumn{1}{c|}{-}           & -        &  & \multicolumn{1}{c|}{V}   & \multicolumn{1}{c|}{340}    & 10  &  & \multicolumn{1}{c|}{V}   & \multicolumn{1}{c|}{186}    & 10  \\
	\multicolumn{11}{l}{\textbf{Note:} ROI = Region of interest; Fix = Fixation; activities are numerically ascending}  \\
\multicolumn{11}{l}{from left to right; S = Start event; E = End event; H = Horizontal; V = Vertical}  
	\end{tabular}
	\label{crazy2}
\end{table}

\begin{table}[]
	\caption{ROIs in process models with hard complexity level}
	\centering
	\scriptsize
\begin{tabular}{ccccccccccc}
	\multicolumn{3}{c}{Straight} &  & \multicolumn{3}{c}{Upwards}                                  &  & \multicolumn{3}{c}{Downwards}                                \\ \cline{1-3} \cline{5-7} \cline{9-11} 
	\multicolumn{3}{c}{Process Model 5}                                         &  & \multicolumn{3}{c}{Process Model 11}                         &  & \multicolumn{3}{c}{Process Model 17}                         \\ \cline{1-3} \cline{5-7} \cline{9-11} 
	\multicolumn{1}{c|}{ROI}      & \multicolumn{1}{c|}{No Fix}      & Hit      &  & \multicolumn{1}{c|}{ROI} & \multicolumn{1}{c|}{No Fix} & Hit &  & \multicolumn{1}{c|}{ROI} & \multicolumn{1}{c|}{No Fix} & Hit \\ \cline{1-3} \cline{5-7} \cline{9-11} 
	\multicolumn{1}{c|}{S}        & \multicolumn{1}{c|}{17}          & 4        &  & \multicolumn{1}{c|}{S}   & \multicolumn{1}{c|}{15}     & 4   &  & \multicolumn{1}{c|}{S}   & \multicolumn{1}{c|}{3}      & 3   \\
	\multicolumn{1}{c|}{1}        & \multicolumn{1}{c|}{80}          & 9        &  & \multicolumn{1}{c|}{1}   & \multicolumn{1}{c|}{74}     & 10  &  & \multicolumn{1}{c|}{1}   & \multicolumn{1}{c|}{47}     & 10  \\
	\multicolumn{1}{c|}{2}        & \multicolumn{1}{c|}{100}         & 9        &  & \multicolumn{1}{c|}{2}   & \multicolumn{1}{c|}{97}     & 10  &  & \multicolumn{1}{c|}{2}   & \multicolumn{1}{c|}{87}     & 10  \\
	\multicolumn{1}{c|}{3}        & \multicolumn{1}{c|}{85}          & 9        &  & \multicolumn{1}{c|}{3}   & \multicolumn{1}{c|}{72}     & 10  &  & \multicolumn{1}{c|}{3}   & \multicolumn{1}{c|}{55}     & 10  \\
	\multicolumn{1}{c|}{4}        & \multicolumn{1}{c|}{82}          & 9        &  & \multicolumn{1}{c|}{4}   & \multicolumn{1}{c|}{60}     & 10  &  & \multicolumn{1}{c|}{4}   & \multicolumn{1}{c|}{50}     & 10  \\
	\multicolumn{1}{c|}{5}        & \multicolumn{1}{c|}{60}          & 9        &  & \multicolumn{1}{c|}{5}   & \multicolumn{1}{c|}{83}     & 10  &  & \multicolumn{1}{c|}{5}   & \multicolumn{1}{c|}{41}     & 10  \\
	\multicolumn{1}{c|}{6}        & \multicolumn{1}{c|}{43}          & 9        &  & \multicolumn{1}{c|}{6}   & \multicolumn{1}{c|}{76}     & 10  &  & \multicolumn{1}{c|}{6}   & \multicolumn{1}{c|}{37}     & 9   \\
	\multicolumn{1}{c|}{7}        & \multicolumn{1}{c|}{47}          & 9        &  & \multicolumn{1}{c|}{7}   & \multicolumn{1}{c|}{85}     & 10  &  & \multicolumn{1}{c|}{7}   & \multicolumn{1}{c|}{41}     & 9   \\
	\multicolumn{1}{c|}{8}        & \multicolumn{1}{c|}{48}          & 9        &  & \multicolumn{1}{c|}{8}   & \multicolumn{1}{c|}{69}     & 10  &  & \multicolumn{1}{c|}{8}   & \multicolumn{1}{c|}{41}     & 10  \\
	\multicolumn{1}{c|}{9}        & \multicolumn{1}{c|}{36}          & 9        &  & \multicolumn{1}{c|}{9}   & \multicolumn{1}{c|}{58}     & 10  &  & \multicolumn{1}{c|}{9}   & \multicolumn{1}{c|}{34}     & 10  \\
	\multicolumn{1}{c|}{E}        & \multicolumn{1}{c|}{14}          & 3        &  & \multicolumn{1}{c|}{E}   & \multicolumn{1}{c|}{15}     & 5   &  & \multicolumn{1}{c|}{E}   & \multicolumn{1}{c|}{2}      & 1   \\
	\multicolumn{1}{c|}{-}        & \multicolumn{1}{c|}{-}           & -         &  & \multicolumn{1}{c|}{H}   & \multicolumn{1}{c|}{431}    & 10  &  & \multicolumn{1}{c|}{H}   & \multicolumn{1}{c|}{308}    & 10  \\
	\multicolumn{1}{c|}{-}        & \multicolumn{1}{c|}{-}           & -        &  & \multicolumn{1}{c|}{V}   & \multicolumn{1}{c|}{412}    & 10  &  & \multicolumn{1}{c|}{V}   & \multicolumn{1}{c|}{210}    & 10  \\
	&                                  &          &  &                          &                             &     &  &                          &                             &     \\
	\multicolumn{3}{c}{Process Model 6}                                         &  & \multicolumn{3}{c}{Process Model 12}                         &  & \multicolumn{3}{c}{Process Model 18}                         \\ \cline{1-3} \cline{5-7} \cline{9-11} 
	\multicolumn{1}{c|}{ROI}      & \multicolumn{1}{c|}{No Fix}      & Hit      &  & \multicolumn{1}{c|}{ROI} & \multicolumn{1}{c|}{No Fix} & Hit &  & \multicolumn{1}{c|}{ROI} & \multicolumn{1}{c|}{No Fix} & Hit \\ \cline{1-3} \cline{5-7} \cline{9-11} 
	\multicolumn{1}{c|}{S}        & \multicolumn{1}{c|}{8}           & 1        &  & \multicolumn{1}{c|}{S}   & \multicolumn{1}{c|}{11}     & 3   &  & \multicolumn{1}{c|}{S}   & \multicolumn{1}{c|}{0}      & 0   \\
	\multicolumn{1}{c|}{1}        & \multicolumn{1}{c|}{56}          & 10       &  & \multicolumn{1}{c|}{1}   & \multicolumn{1}{c|}{95}     & 9   &  & \multicolumn{1}{c|}{1}   & \multicolumn{1}{c|}{63}     & 10  \\
	\multicolumn{1}{c|}{2}        & \multicolumn{1}{c|}{102}         & 10       &  & \multicolumn{1}{c|}{2}   & \multicolumn{1}{c|}{115}    & 9   &  & \multicolumn{1}{c|}{2}   & \multicolumn{1}{c|}{110}    & 10  \\
	\multicolumn{1}{c|}{3}        & \multicolumn{1}{c|}{96}          & 10       &  & \multicolumn{1}{c|}{3}   & \multicolumn{1}{c|}{97}     & 8   &  & \multicolumn{1}{c|}{3}   & \multicolumn{1}{c|}{83}     & 9   \\
	\multicolumn{1}{c|}{4}        & \multicolumn{1}{c|}{76}          & 10       &  & \multicolumn{1}{c|}{4}   & \multicolumn{1}{c|}{79}     & 8   &  & \multicolumn{1}{c|}{4}   & \multicolumn{1}{c|}{63}     & 10  \\
	\multicolumn{1}{c|}{5}        & \multicolumn{1}{c|}{63}          & 10       &  & \multicolumn{1}{c|}{5}   & \multicolumn{1}{c|}{71}     & 8   &  & \multicolumn{1}{c|}{5}   & \multicolumn{1}{c|}{51}     & 10  \\
	\multicolumn{1}{c|}{6}        & \multicolumn{1}{c|}{63}          & 10       &  & \multicolumn{1}{c|}{6}   & \multicolumn{1}{c|}{72}     & 9   &  & \multicolumn{1}{c|}{6}   & \multicolumn{1}{c|}{46}     & 10  \\
	\multicolumn{1}{c|}{7}        & \multicolumn{1}{c|}{63}          & 9        &  & \multicolumn{1}{c|}{7}   & \multicolumn{1}{c|}{69}     & 9   &  & \multicolumn{1}{c|}{7}   & \multicolumn{1}{c|}{54}     & 8   \\
	\multicolumn{1}{c|}{8}        & \multicolumn{1}{c|}{47}          & 10       &  & \multicolumn{1}{c|}{8}   & \multicolumn{1}{c|}{86}     & 9   &  & \multicolumn{1}{c|}{8}   & \multicolumn{1}{c|}{52}     & 10  \\
	\multicolumn{1}{c|}{9}        & \multicolumn{1}{c|}{54}          & 10       &  & \multicolumn{1}{c|}{9}   & \multicolumn{1}{c|}{107}     & 9   &  & \multicolumn{1}{c|}{9}   & \multicolumn{1}{c|}{54}     & 10  \\
	\multicolumn{1}{c|}{10}       & \multicolumn{1}{c|}{59}          & 10       &  & \multicolumn{1}{c|}{10}  & \multicolumn{1}{c|}{92}     & 8   &  & \multicolumn{1}{c|}{10}  & \multicolumn{1}{c|}{52}     & 10  \\
	\multicolumn{1}{c|}{11}       & \multicolumn{1}{c|}{43}          & 9        &  & \multicolumn{1}{c|}{11}  & \multicolumn{1}{c|}{77}     & 9   &  & \multicolumn{1}{c|}{11}  & \multicolumn{1}{c|}{32}     & 9   \\
	\multicolumn{1}{c|}{E}        & \multicolumn{1}{c|}{11}          & 3        &  & \multicolumn{1}{c|}{E}   & \multicolumn{1}{c|}{9}      & 4   &  & \multicolumn{1}{c|}{E}   & \multicolumn{1}{c|}{1}      & 1   \\
	\multicolumn{1}{c|}{-}        & \multicolumn{1}{c|}{-}           & -         &  & \multicolumn{1}{c|}{H}   & \multicolumn{1}{c|}{567}    & 9   &  & \multicolumn{1}{c|}{H}   & \multicolumn{1}{c|}{442}    & 10  \\
	\multicolumn{1}{c|}{-}        & \multicolumn{1}{c|}{-}           & -         &  & \multicolumn{1}{c|}{V}   & \multicolumn{1}{c|}{531}    & 9   &  & \multicolumn{1}{c|}{V}   & \multicolumn{1}{c|}{313}    & 10 \\
	\multicolumn{11}{l}{\textbf{Note:} ROI = Region of interest; Fix = Fixation; activities are numerically ascending}  \\
	\multicolumn{11}{l}{from left to right; S = Start event; E = End event; H = Horizontal; V = Vertical}  
\end{tabular}
	\label{crazy3}
\end{table}

\begin{figure*}[hb]
	\centering
	\includegraphics[width=\linewidth]{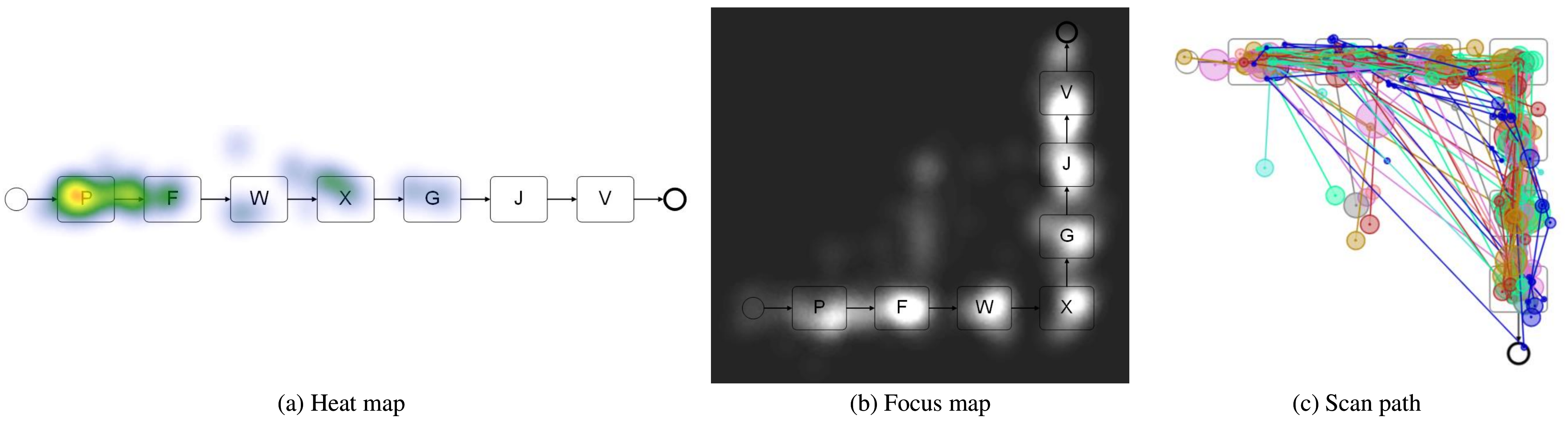}
	\caption{Visual analysis of recorded eye movements showing superimposed (a) heat map on the straight mapping, (b) focus map on the upward mapping, and (c) scan paths on the downward mapping}
	\label{analysis}
\end{figure*}

\subsection{Gaze Patterns}
\label{visual pattern}

Based on obtained results and to better understand participants' attention and scanning behavior, SMI BeGaze was used for the analysis of recorded eye movements with additional visualization techniques. More specifically, eye movements were visualized in terms of gaze plots (e.g., scan paths) and attention maps (e.g., heat maps) in order to determine whether specific gaze patterns were deployed during process model comprehension. The Figures~\ref{analysis} (a) - (c) present the used techniques for the visual analysis of the eye movements. In particular, Figure~\ref{analysis} (a) shows a superimposed heat map (i.e., focus of visual attention) from all participants on process model 4 (i.e., straight mapping, medium level of complexity; see Table \ref{models}) after one second of model comprehension. It can be seen from the figure that the fixation density is highest at the first activity of the model. In contrast to a heat map, a focus map (i.e., alteration of the luminescence of the process model based on the amount of received visual attention) is depicted in Figure~\ref{analysis} (b) on process model 10 (i.e., upward mapping, medium level of complexity; see Table \ref{models}) after a first comprehension iteration (i.e., each activity was fixated once, from the beginning to the end of the model). From the focus map, it is recognizable that participants focused mostly on just the activities, which have been fixated at least once during the first comprehension iteration. Moreover, the performed eye movements were oriented along with the process flow of the model (i.e., horizontal to the left until the kink and from there further upwards). Finally, Figure~\ref{analysis} (c) presents plotted scan paths (i.e., chronological order of fixations and saccades) from the participants after finishing the comprehension of process model 16 (i.e., downward mapping, medium level of complexity; see Table \ref{models}). It is discernible that eye movements were following the mapping structure, similar as presented in Figure~\ref{analysis} (b). In addition, eye movements reflected back-and-forth saccade jumps altering between activities in the process model. Thereby, these jumps went backwards, forwards, but also diagonally, as shown in the figure. In general, the illustrated eye movements presented in Figures~\ref{analysis} (a) - (c) can be identified in most of the used process models in the study. 

On the basis of the obtained results and the performed visual analyses (see Figures~\ref{analysis} (a) - (c)) in the context of process model literacy, three distinct gaze patterns have been identified from the data, which reflect common eye movements performed by the participants during the study task (i.e., process model comprehension): orientation, comprehension, and congruence pattern. Note that there exists no clear separation between the gaze patterns and a change between the pattern is based on an intertwined transition.

\begin{figure*}[ht]
	\centering
	\includegraphics[width=\linewidth]{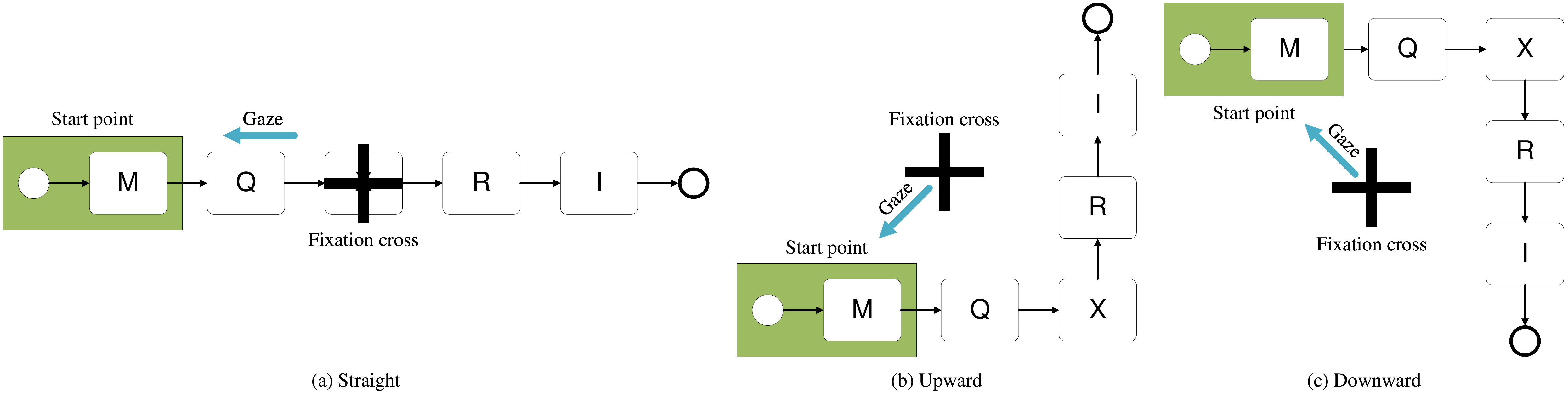}
	\caption{Orientation pattern for process model mapping (a) straight, (b) upward, and (c) downward}
	\label{op}
\end{figure*}

\subsubsection{\textbf{Orientation pattern}}

The first gaze pattern (i.e., orientation pattern) occurred at the beginning of the comprehension. More specifically, attention was fixated around the center of the display through the presented fixation cross. After the disappearance of the fixation cross and the subsequent presentation of the process model, the participants tried to locate the start of the model. The participants first assumed intuitively (e.g., culture-dependent reading from left to right) and then learned as the study progressed, that the start of the model was always presented on the left side of the monitor. Depending on the mapping type, the start of the model was either on the center left (i.e., straight), bottom left (i.e., upward), or top left (i.e., downward). Hence, the gaze shifted quickly towards the start of the model, which constituted either the start event or mostly the first activity after the start event in the model. The eye movements showed similarities to the described elemental operations indexing of salient elements or visual search (see Section~\ref{theo}). The start of the process models constituted an anchor, from which the participants started with further attention processing (i.e., comprehension). The Figures~\ref{op} (a) - (c) illustrate the orientation pattern in the three process models mappings (i.e., (a) straight, (b) upward, (c) downward). Thereby, gaze direction from the fixation cross as well as the start point as ROI is shown in the figures. As described in Section~\ref{eye}, the identification of the start was made by either a direct or indirect fixation. However, a few participants skimmed through the process models at first. Thereby, skimming (e.g., shorter fixation durations) is a rapid reading pattern in order to get - inter alia - a general overview of the material (e.g., structure) \cite{duggan2009text}. The occurrence of skimming was seen more frequently in process models with an increased level of complexity. Thereby, it was observed that the participants identified the end of the process model first and skimmed through the model backwards to the start point. A reason for the use of skimming is that participants wanted to get a first impression of what to expect (e.g., identification of the mapping and level of complexity). Thereby, it seems that the use of skimming reflected the elemental operation boundary tracing. Moreover, characteristic in this pattern was the appearance of long saccades (i.e., especially in the complex process models), which stabilized with first fixations on the start point of the process model. The appearance of the orientation pattern was short-lived (i.e., max 1 second at most), as participants were able to identify the start of the process models quickly. Thereby, depending on the level of complexity, the process models were shown near the previously displayed fixation cross, which eased the fixation on the start of the model (e.g., easy process models with a straight mapping appeared directly in the center of the monitor).   

\subsubsection{\textbf{Comprehension pattern}}

\begin{figure*}[ht]
	\centering
	\includegraphics[width=\linewidth]{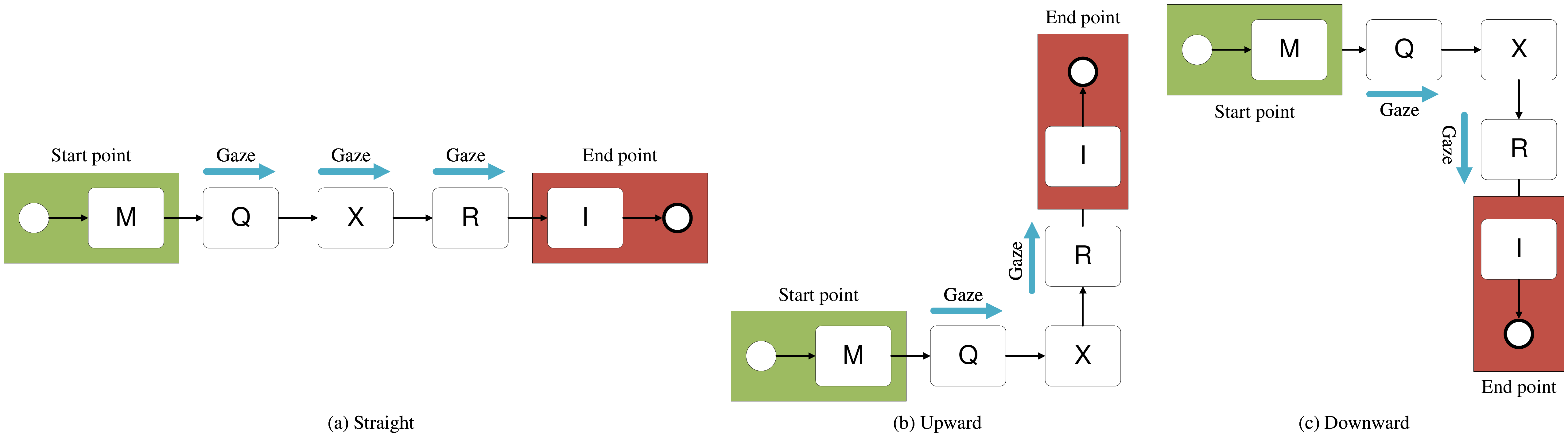}
	\caption{Comprehension pattern for process model mapping (a) straight, (b) upward, and (c) downward}
	\label{cp}
\end{figure*}

After the orientation pattern was completed, the subsequent comprehension pattern emerged. More specifically, after the identification of the start point of the presented process model, the participants usually fixated the activities in the order represented by the process flow until the end point of the model. Hence, eye movements of the participants reflected the mapping structure of the presented model (see Figure~\ref{analysis} (b)). The Figures~\ref{cp} (a) - (c) depict the comprehension pattern in the three mappings (i.e., (a) straight, (b) upward, (c) downward). Further, the figures are showing the gaze direction from the two ROIs: start and end point. Fixations were performed in the scope of the model boundaries indicating similarities to the elemental operation of boundary tracing (i.e., the gaze was within the process model contours). Thereby, the duration for this pattern (i.e., usually between 3 and 10 seconds) depended on process model complexity and mapping (e.g., shorter duration in straight mappings). Juxtaposed to the orientation pattern, saccades and related amplitudes were shorter, while the average fixation duration was higher (i.e., typical duration for silent reading \cite{liversedge2006binocular}). In this pattern, the first extraction of semantic information from the activity labels took place, and we hypothesize (see ~\ref{future}), as known from literature \cite{o1992comprehension}, that a mental model (i.e., internal representation of external reality) of the shown process model was constructed in this pattern. Furthermore, with increasing level of model complexity, it was observed that some activities (i.e., activity in the middle between two activities) were skipped and, hence, not fixated. Due to this observation, the question arises whether the information of the skipped activities was obtained with an indirect fixation (i.e., peripheral vision), suppressed, or was marked as congruence point for a later return (see Section~\ref{future}). Eye movements were not only focused forward along with the process flow, but also regressive eye movements (i.e., regressions) were discernible between abreast activities. Thereby, a regression constitutes a backward jump of the attention to a prior fixation point. Moreover, regressions are important features in order to maintain attention allowing for a correction of the perception made before \cite{eskenazi2017regressions}. As the number of activities increased, so did the number of regressions. Thus, the regressed activities were marked as a congruence points for further processing. We assume that the appearance of regressions was the result of the non-lexicographic labeling of the activities. Moreover, participants tended to form activity pairs (i.e., chunks) in their comprehension strategies, as chunks of information can be memorized more easily, juxtaposed to atomic information \cite{thalmann2019does}. In this context, as seen in the results presented in Section~\ref{eye}, participants divided the two sequences (i.e., horizontal and vertical) from the upward and downward mapping into chunks as well. This division strategy may be referred to a visual routine, in which elemental operations (e.g., delimited activation) are concerned with the separation of a monolithic structure (i.e., process model) into separated units (i.e., horizontal and vertical). Based on this division, first the horizontal sequence was considered and, afterwards, the vertical. 

\subsubsection{\textbf{Congruence pattern}}

\begin{figure*}[ht]
	\centering
	\includegraphics[width=\linewidth]{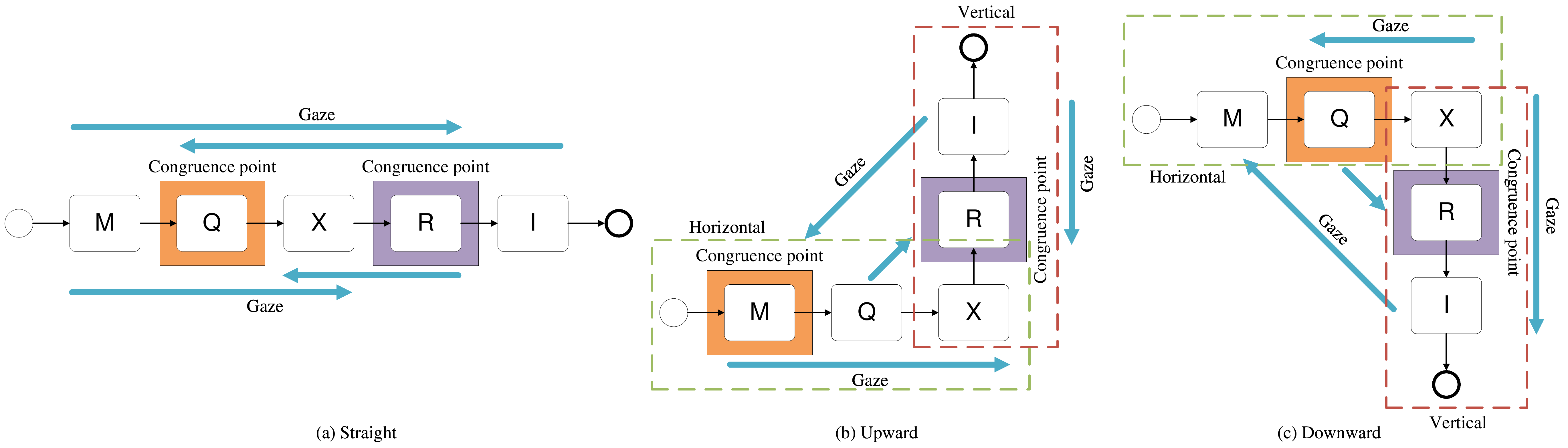}
	\caption{Congruence pattern for process model mapping (a) straight, (b) upward, and (c) downward}
	\label{cop}
\end{figure*}

Finally, the congruence pattern usually appeared after the first comprehension iteration (i.e., comprehension pattern; each activity was fixated once from the beginning to the end of the process model) and was presumably used to facilitate attention during the comprehension of the presented process model. More specifically, we suggest that the created mental model from the comprehension pattern was checked for congruency and prior processed information was reconciled. Hence, the attention (i.e., gaze) was shifting, usually from the endpoint of the model (i.e., where the comprehension pattern ended), towards specific congruence points. A congruence point represents either a single activity or a chunk of activities, in which the gaze was alternating. Prior marked points in the process models were considered again for the reprocessing of contained information. However, several participants started the comprehension anew in the presented process model, with an increased fixation number as well as duration on the activities. Thereby, the eye movements were not bound to the process model mapping, but diagonal saccades were identified (i.e., superimposed scan paths created a triangle shape on the model). Thereby, the comprehension of a process model in the congruence pattern led to the application of different elemental operations, which are composed into visual routines. For example, a conceivable operation describes that in the upward and downward mapping a mnemonic (i.e., technique for information retention \cite{dresler2017mnemonic}) was constructed, which connected the respective two sequences (i.e., horizontal and vertical). In contrast, this observation was not seen in the straight mapping and attention moved back and forth on a horizontal level. The Figures~\ref{cop} (a) - (c) outline the congruence pattern in the three mappings (i.e., (a) straight, (b) upward, (c) downward). Moreover, potential gaze directions as well as congruence points as ROIs are illustrated in the figures. Thereby, shifts in the attention (e.g., fixations, related durations, and saccades) differed substantially within this pattern  (i.e., longer vs. shorter). A reason can be that this pattern was more dependent on person-related characteristics. For example, we assume that the duration of this pattern was shorter in an individual with a high degree of process model literacy and well developed memorization capabilities (see Section~\ref{future}). Furthermore, from the properties of fixations, rapid reading patterns were identified, which represented a scanning rather than a skimming pattern. Contrary to skimming (i.e., as seen in the orientation pattern), scanning is used for a more fine-grained information processing (i.e., longer fixation duration). As described in the comprehension pattern, participants, who applied the approach to divide the two sequences into chunks (i.e., horizontal as well as vertical and a first comprehension thereof, consequently spent less time in the congruence pattern. Other participants, in turn, applied respective strategy in the congruence pattern. Thereby, the number of regressions was higher compared to the comprehension pattern. Regarding the described saturation effect (see Section~\ref{eye}), it was apparent, especially in the straight as well as downward mapping (i.e., vertical sequence), how the number of fixations and related decreased in the course of the process flow. However, fixations reflected an even distribution in the upward mapping. 

\subsection{Visual Routines During Process Model Comprehension}

In graph literacy, the comprehension of a graph (e.g., a diagram explaining the compression of carbon dioxide \cite{pisan1995visual}) with inherent properties (e.g., spatial relations, data points) proceeds in the visual system, respective cognitive control, according to the framework proposed in \cite{ullman1987visual} with visual routines and elemental operations, of which these routines are composed of (see Section~\ref{theo}; \cite{michal2016visual}). Thereby, the deployment of such routines and operations (i.e., attention shift) are controlled by the cognitive control \cite{ballard1995computational}. Depending on the objective and the conditions, the cognitive control applies visual routines, which are necessary for the achievement of the current objective (e.g., make a peanut butter and jelly sandwich \cite{hayhoe2005eye}). In the context of process model literacy, confirmed by our performed analyses of the recorded eye movements from this exploratory study (i.e., gaze patterns; see Section~\ref{visual pattern}), it appears that process model comprehension results in the exhibition of various visual routines and associated elemental operations. Thereby, the proper comprehension of the process model and the correct answering of related comprehension questions represented the overarching objective in the study. Hence, based on the active gaze pattern, the cognitive control applies visual routines, which are required in respective pattern in order to reach the specific objective. For example, in the orientation pattern, it was important for participants to identify the start of the presented process model. The start constituted the process model element that was either represented by the start event or the first activity of the model. As a result, the participants were engaged in a visual routine referring, for example, to element detection. The respective element (i.e., start of the process model) possessed the geometric shape similar to a circle (i.e., event) or a rectangle (i.e., activity). Hence, for this kind of visual routine, specific elemental operations (e.g., visual search and indexing of salient elements; see Section~\ref{theo}) were necessary that are concerned with the identification of the appropriate shape. On the basis of this deliberation and the results obtained, the following Figure~\ref{theoModel} illustrates a defined model regarding the deployment of visual routines and elemental operations during process model comprehension. With the cognitive control as operative (e.g., for attention deployment) and based on the active gaze pattern, a specific visual routine is chosen in order to achieve the current objective (e.g., identify start of the process model). The chosen visual routine, in turn, is composed of a set of elemental operations. The latter summarizes operations that are mandatory to extract relevant information from a process model. They are dependent on the following process model properties shape, structure, and content, which are defined as follows:

\smallskip

\begin{itemize}
	\item Shape: Summarizes elemental operations that are concerned with the identification of the geometric shape of the elements in a process model, i.e., circle = event, rectangle = activity.
	\item Structure: Refers to the structural representation (i.e., mapping) of a process model, i.e., sequence, upward, downward.
	\item Content: Describes the associated content of the process model elements, i.e., labels of activities.
\end{itemize}

\smallskip

Vice versa, the cognitive control receives a response from selected routine when respective information is found or not (i.e., completed objective), since the relationship between the aspects (i.e., operative, pattern, routine, operation) is bidirectional. Following this, the cognitive control is able to evoke the next visual routine based on the received response (e.g., when an activity has been detected, the label can be read). In the context of the conducted study, the following five visual routines could be identified that are necessary for a proper process model comprehension in this context:

\smallskip

\begin{itemize}
	\item \raisebox{.5pt}{\textcircled{\raisebox{-.9pt} {1}}} Element Detection: This routine is concerned with the correct identification of respective elements in a process model and to distinguish them with each other. 
	\item \raisebox{.5pt}{\textcircled{\raisebox{-.9pt} {2}}} Label Reading: The contained information in the specific model elements (i.e., labels) is extracted and processed accordingly. 
	\item \raisebox{.5pt}{\textcircled{\raisebox{-.9pt} {3}}} Flow Following: The gaze (i.e., attention shift) is oriented along the process flow and in the scope of respective mapping type.
	\item \raisebox{.5pt}{\textcircled{\raisebox{-.9pt} {4}}} Partition: The partition of a monolithic structure into smaller chunks (i.e., vertical and horizontal) is described in this visual routine.
	\item \raisebox{.5pt}{\textcircled{\raisebox{-.9pt} {5}}} Reconciliation: Existing or recently recorded information in the working memory load is verified and checked for correctness.
\end{itemize}

\smallskip

It appears that in the orientation pattern solely the visual routine \raisebox{.5pt}{\textcircled{\raisebox{-.9pt} {1}}} Element Detection is of importance for the identification of the start of the presented process model. No semantic, syntactic or structural information from the process model is obtained in this pattern.  Whereas in the comprehension pattern nearly all routines \raisebox{.5pt}{\textcircled{\raisebox{-.9pt} {1}}} - \raisebox{.5pt}{\textcircled{\raisebox{-.9pt} {4}}} are invoked. Based on the objectives given by the cognitive control, the contained semantic information from the single activities in a process model are read and processed. Moreover, based on the insights obtained, the partition of the process model into chunks is performed in the comprehension pattern as well. This routine is only relevant in the down or upward mapping.  Finally, in the congruence pattern, the visual routines \raisebox{.5pt}{\textcircled{\raisebox{-.9pt} {2}}}, \raisebox{.5pt}{\textcircled{\raisebox{-.9pt} {3}}}, and \raisebox{.5pt}{\textcircled{\raisebox{-.9pt} {5}}} interact with the cognitive control. We assume that most information from the presented process model is extracted in this pattern. The other two visual routines \raisebox{.5pt}{\textcircled{\raisebox{-.9pt} {1}}} and \raisebox{.5pt}{\textcircled{\raisebox{-.9pt} {4}}} are not invoked, since the information was already stored in the working memory by the execution of respective visual routines in the preceding patterns. In addition, the visual routine \raisebox{.5pt}{\textcircled{\raisebox{-.9pt} {5}}} Reconciliation is exclusively invoked in this pattern. The latter is used to validate present (i.e., stored) or marked (i.e., anchor) information in a process model stored in the working memory. \raisebox{.5pt}{\textcircled{\raisebox{-.9pt} {5}}} Reconciliation is mainly used as a preparation for the overarching objective (i.e., answering of the comprehension questions). Note that the current model summarizes the first general observation and a specialization (e.g., non-corresponding behavior seen in respective pattern) needs to be investigated in future work (see Section~\ref{future}). Generally, the model presented in Figure~\ref{theoModel} is in line with related work known from other domains (e.g., \cite{ballard2009modelling,cavanagh2001attention}).

Information processing of the visual system results into a set of different visual routines, which are pieced together by the cognitive control with other aspects (e.g., person-related characteristics), thus representing the process of process model comprehension. On the basis of the presented model shown in Figure~\ref{theoModel}, the questions arise, which additional visual routines are invoked during process model comprehension and of which elemental operations are these routines composed of? The investigation of such questions may be promising directions for further research in order to get a better understanding of working with process models (see Section~\ref{future}). For example, knowing that the elemental operation visual search is frequently applied, process models or model viewer tools may be designed in such ways that this operation is facilitated.

\subsection{Discussion}
\label{discussion}

\begin{figure}[t]
	\centering
	\includegraphics[width=\linewidth]{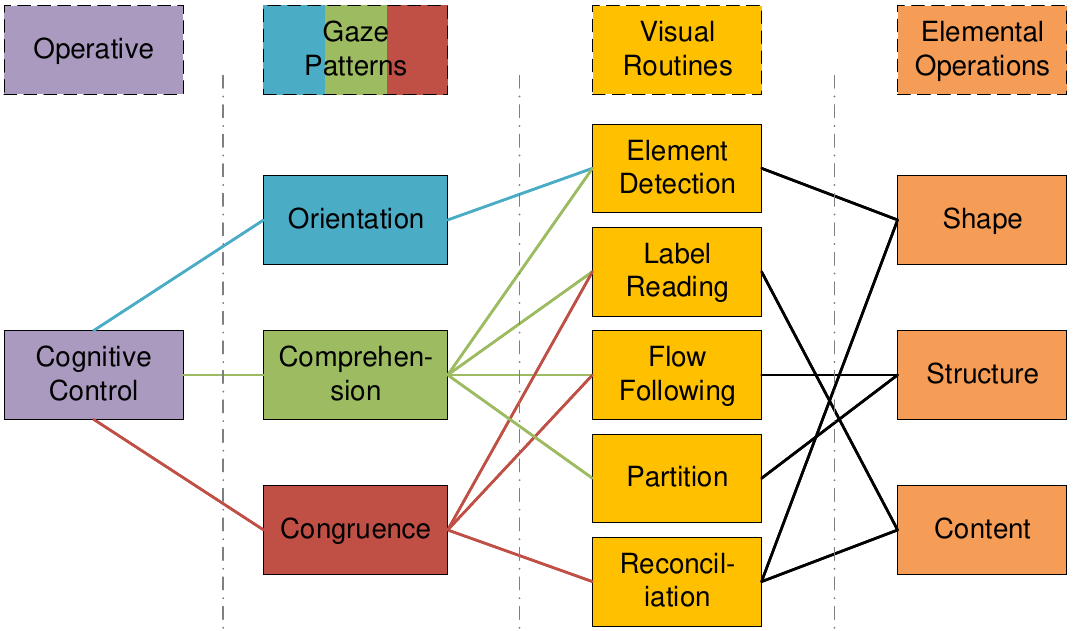}
	\caption{Model for the deployment of visual routines and elemental operations during process model comprehension}
	\label{theoModel}
\end{figure}

In general, as seen in the descriptive statistics (see Section~\ref{descriptive}), there was an increase in the performance measures with an increasing level of process model complexity. In more detail, participants needed more time for the comprehension of the presented models, performed more eye movements, achieved fewer points in the comprehension questions and, in addition, needed more time for answering respective questions. In general, such a result was expected and has been observed in related work in this context (e.g., \cite{tallon2019comprehension,sanchez2017case}). With an increasing number of activities, more information must be captured as well as processed, thus having an influence on respective performance measures. Furthermore, as cognitive resources are limited, there is a raised probability for errors (e.g., model misinterpretation) \cite{oberauer2016limits}. Since the process models used in the study were evenly distributed regarding model complexity (i.e., the easiest model with 1, the most difficult with 11 activities; adding 2 activities each per complexity level; see Section~\ref{material}), no correlation between level of complexity and defined performance measures was identified. More specifically, minor changes were observed in the performance measures first, followed by a distinct increase therein. However, variances in the performance measures were collected for the individual mapping types (i.e., straight, upward, downward; see Section~\ref{material}). For example, regarding the downward mapping, to the penultimate process model, a steady increase in the duration was discernible, which increased conspicuously in the last model (see Table~\ref{descriptive}). The inferential statistics (see Section~\ref{inferential}) confirmed the differences along with the level of complexity, but, in addition, significant differences between the process model mappings were identified as well (e.g., score, fixation; see Table~\ref{inf}). Regarding the score presented in Table~\ref{cqdscr}, it can be determined that the comprehension questions referring to the upward process model mapping were less correctly answered, juxtaposed to the other two mapping types. In turn, response times for answering respective questions did not differ substantially. Further, the mapping preference shown in Table~\ref{cltdscr} underlines this finding. To be more precise, participants indicated that the straight mapping (n = 18) was perceived as the most, whereas the upward mapping (n = 3) constituted the least comprehensible process model mapping. An explanation may be that the reading direction in the upward mapping (i.e., vertical bottom-up process flow) represented an atypical direction according to respective literature \cite{gobel2015up}. Similar results in process model comprehension have been reported by the authors in \cite{figl2014importance}. In turn, attention of the participants was conveyed more effectively in the comprehension of process models with a downward mapping (e.g., fewer fixations with higher comprehension scores). The visual system extracted and processed the presented information in the process models in a more effective way. A rationale, as shown in literature \cite{hegarty2011cognitive}, describes the spatial closeness of the elements in the model (i.e., scope of attention is smaller). Based on the descriptive as well as inferential statistics, we assume that the differences between the model mappings (i.e., especially in upward type) would be more striker in larger study samples (see Section~\ref{limit}). In this regard, dimensions related to the cognitive load (i.e., intrinsic, extraneous, germane) were at an intermediate level (i.e., 3.5 with max 7; see Table~\ref{cltdscr}). Accordingly, participants coped with challenges concerning the complexity of the information presented in the process models (e.g., mapping, labeling). As described in \cite{moody2009physics}, retinal (e.g., size) and planar (e.g., structure) variables for constructing a process model exist, which are affecting the cognitive load. Hence, the upward process model mapping required more cognitive effort for proper model comprehension. This is an interesting finding as process models from practice usually consist of a composition of different process flow directions \cite{weber2011refactoring}. Thereby, as an implication it is advisable to avoid upward process flows in the creation of such models and the recommendation to use a downward mapping. Furthermore, the findings concerning the cognitive load confirmed that process model comprehension, in general, constitutes a complex matter. Juxtaposed to process models applied in the real world, the process models in this study were of a simple nature and composed only basic elements from BPMN 2.0 (i.e., events, activities, sequence flow), with a maximum number of 11 activities. However, these simple process models have already created a moderate effect on the cognitive load. As a consequence, for organizations applying process models, it is important to note that even less complex process models may be challenging regarding their proper comprehension. Therefore, an emphasis should be put into approaches (e.g., modularization \cite{winter2021measuring}) referring to complexity reduction (e.g., size) in such models (i.e., a recommended benchmark constitutes 50 elements \cite{mendling2010seven}). Moreover, participants tended to positively perceive the application of process models in practice and, further, were of the opinion that such models can foster the comprehension of processes and related information (see~Table \ref{cltdscr}). Thereby, attention should be paid to a reduced cognitive load with, for example, the provision of proper explanations about the semantic as well as syntactic information in a process model \cite{wang2017effect}. 

The further analyses of the attention revealed that during the comprehension of the process models learning effects as well as strategies emerged. For example, the left-justified orientation of the gaze at the beginning of the comprehension or, as can be noticed in the Tables~\ref{crazy1} - \ref{crazy3}, the events (i.e., start and end) in the models were considered only by a few participants. A reason suggests that the events did not contain any semantic information (i.e., no label description) and, hence, did not reflect any useful information. As a result, visual capacities did not have to be placed on the events (i.e., suppressed) and can be used otherwise (e.g., on activities). Similar observations have been found in related work \cite{kembhavi2016diagram}, showing that information without added value is likely to be not perceived. On the basis of this insight, it is important to pay attention that all relevant elements in a process models are properly labeled. Otherwise, there is a growing risk that relevant information is not captured, which consequently may lead to an ambiguous comprehension of the model. In the context of the upward and downward mapping, participants tended to split the horizontal and vertical sequence into pairs of groups (i.e., chunks). The individual groups were considered as a closed unit, in which the activities have been comprehended separately, apart from the others. In graph comprehension, due to cognitive capacity limitations, the construction of chunks depicts an important feature regarding the association process between presented visual information to quantitative referents \cite{shah1999graphs}.  On the basis of this observation, we imply the confirmation that forming chunks constitutes an essential factor for the comprehension of process models. In order to preserve a positive effect on the comprehensibility, research suggests creating blocks in the modeling of processes, as such block-structured models are less error-prone \cite{la2011managing}. Moreover, of interest was the observation of a saturation effect in the straight and downward mapping (see Table~\ref{crazy1} and \ref{crazy3}). In more detail, the number of fixations on the activities initially increased and then, from the middle of the process model, decreased in the subsequent activities. The process models did not present new planar information and, as a reason, visual processing (i.e., attention) had to focus solely on the labels of the activities. Related research obtained similar results demonstrating that attention in graph literacy is more directed towards novel and key information \cite{okan2016people}. In turn, this effect was not found in the upward mapping. Contrary to the straight and downward mapping, comprehension was additionally confronted with an atypical process flow direction in the upward mapping. This atypical direction constituted new planar information, leading to an equal distribution of fixations. This observation may indicate that attention can decline not only over time but also along the information presented in a model. For practice, it is advisable to resort on techniques that maintain attention (e.g., coloring \cite{djurica2020impact}) during process model comprehension. 

On the basis of the results and additional visual analyses, three distinct gaze patterns (i.e., orientation, comprehension, congruence) were identified, which describe general patterns for attention deployment during process model comprehension. Thereby, each pattern presents a specific procedure in order to cope with the study objective. Moreover, a model (see Figure~\ref{theoModel}) regarding the deployment of visual routines during process model comprehension was presented. In more detail, depending on the active pattern, the cognitive control initiates the execution of appropriate visual routines to achieve the current (e.g., identification of the start of the process model) as well as long-term objectives (i.e., proper comprehension of the presented process model). This model regarding visual routine during process model comprehension is in compliance with the insights presented in similar research (e.g., \cite{710820,hayhoe2005eye}). Based on the proposal given in \cite{ullman1987visual}, the visual routines are composed of elemental operations that are crucial to extract relevant information from a process model. In general, visual routines and respective operations are important means to be able to better explain information extraction and processing from process models via attention deployment, thus allowing the definition of directives fostering model comprehension. For example, the start point of a process model was assumed to be an anchor, from which further processing continued. Thereby, related research demonstrated that the provision of anchors is decisive for a proper comprehension of graphs and diagrams \cite{huestegge2018integration}. As an implication, such anchors should be defined during process model comprehension (e.g., coloring of the start point \cite{kummer2021effect}). 

As described in Section~\ref{visual pattern}, visual routines consist of the composition of elemental operations, whereas all performed eye movements on a stimulus (i.e., process model) are made up of a variety of visual routines initiated by the cognitive control. However, no concrete assertion can yet be made about which elemental operations exactly belong to a specific visual routine. Indications are present that, for example, visual routine \raisebox{.5pt}{\textcircled{\raisebox{-.9pt} {1}}} Element Detection (see Figure~\ref{theoModel}) relies on operations such as visual search or indexing of salient elements (see Section~\ref{theo}). Furthermore, the identified five visual routines (see Figure~\ref{theoModel}) for process model comprehension are not exhaustive. To be more precise, the proposed visual routines may be adequate for the proper comprehension of the process models used in the presented study. However, for more complex process models, additional visual routines are required in order to cope with the confronted challenges (e.g., comprehension of concrete semantic information of structural model information such as loops). 

In this regard, another issue concerns the sequence of information processing. Based on our observations, we were unable to determine whether the applied visual routines as well as operations were performed in serial or in parallel with others. More specifically, which of the presented information (e.g., retinal or planar properties such as size and structure) in the process models were processed by the visual system simultaneously and which can only be processed in serial (see Section~\ref{future}). Related research suggests different models regarding information processing \cite{little2018information}. In the context of process model literacy, we assume that most operations were performed in serial (see Section~\ref{future}), since studies demonstrated that comprehension processes are more compromised in the handling of multiple information, juxtaposed to serial processing \cite{ludewig2020influences}. 

\subsection{Limiting Factors}
\label{limit}

The study is confronted with limiting factors that need to be addressed in future studies. (1) For this first exploratory study, the used process models (i.e., regarding mapping and level of complexity) were kept simple intentionally (i.e., max number of activities was 11 and only 1 change in the upward as well as downward mapping in the process flow occurred). However, process models from the real world usually document more complex processes with a larger number of activities and a more ramified structure (i.e., mapping). Therefore, the obtained results constituted constraints regarding generalizability. In this context, (2) labels of the activities contained only alphabetical characters and documented no semantic information. As a consequence, labels documenting semantic information (e.g., concrete process steps such as send order) have a different effect on attention (i.e., eye movements), since additional resources must be invested for the reading and comprehension of the labels. Furthermore, (3) another limitation represents the missing option to look back at the process model while, at the same time, answering related comprehension questions. In particular, the process models must be memorized and, hence, there was a growing risk that given answers were guessed due to wrong or incomplete memorization. In addition, since the pure comprehension of process models (i.e., without any guidance) is uncommon, results may differ in, for example, a targeted search for information in process models. However, in this exploratory study, we wanted to deliberately disclose the approaches for the pure comprehension of BPMN 2.0 process models. (4) The sizes of the samples also limit the statistical power and there might be significant differences between the process model mappings, which we could not reveal in this study, but which might become apparent in larger sample sizes. Further, (5) although only 6 values have been imputed for each performance measure (see Section~\ref{measures}) using Expectation-Maximization (i.e., non-significant Little's MCAR test), results from inferential statistics might have been affected by the applied imputation technique. (6) The used quasi-experimental design with the application of a defined permutation table for the balanced distribution of the process model mappings as well as complexity levels represent another limiting factor. Since no randomization was applied, the ability to conclude a causal association between the process models mappings and the obtained result might be limited. Moreover, (7) the ROIs were drawn conservatively larger and protruded slightly above the activities. Consequently, the risk for error prone ROI-fixation detection was increased \cite{orquin2018threats}. Finally, (8) the study was conducted during the COVID-19 pandemic, in which specific hygiene measures (e.g., spatial distance) as well as regulations (e.g., the permanent wearing of a face mask) had to be strictly followed. As a result, this particular circumstance may have influenced the behavior of the participants. Generally, we justify confronted limitations by the fact that this study was based on an explorative rationale, instead of a study aiming to replicate findings from similar studies. However, while results look promising, the replication or additional studies are needed in order to confirm the generalization of the results (see Section~\ref{future}).

\subsection{Future Work}
\label{future}

The conducted exploratory eye tracking study constitutes the first empirical exploration of visual routines and contributes to the vast body of research in the context of process model literacy \cite{figl2017comprehension}. Moreover, based on the results obtained, a number of promising directions arise for further studies. A continuation of this research delineates a more fine-grained investigation of the proposed model (see Figure~\ref{theoModel}) regarding visual routines and elemental operations during process model comprehension. These include, for example, the identification of additional visual routines while comprehending (more complex) process models as well as the determination of which elemental operations are composed in a specific routine. Therefore, conformity or interference factors (e.g., anticipated vs unanticipated elements) may be juxtaposed in order to observe applied approaches of the visual system (i.e., cognitive control) to cope with these features. Furthermore, are there specific differences (e.g., longer vs. shorter) in attention parameters (e.g., eye movements such as regression, indirect fixation) in process model comprehension during the exhibition of visual routines. This will allow - inter alia - the definition of process model-dependent metrics regarding eye movements (e.g., correlation of fixations with the number of elements in a model) based on retinal and planar properties. In general, the confirmation of validation of the three identified gaze patterns (i.e., orientation, comprehension, and congruence; see Section~\ref{visual pattern}) needs to be done in further studies (e.g., in more complex process models). The findings can, in turn, be used for an automatic detection of the current gaze pattern in a process model viewer tool enabling the provision of a more effective assistance during model comprehension (e.g., the initial coloring of the start event as an anchor in the orientation pattern). This study focused solely on process model comprehension without any concrete task and, hence, an emphasis was put on memorization. Usually, the comprehension of process models is related to a given task (e.g., what are my tasks in the process?), in which respective information needs to be found and extracted from a model. Therefore, it would be interesting to investigate how visual routines are associated during the search for information and whether identified patterns from this work or even novel ones can be found. In this study, the process flow was oriented based on the three process model mapping types (i.e., straight, upward, downward; see Section~\ref{material}), which affected the reading direction of the participants. As can be seen from the results, the upward mapping constituted a greater challenge during process model comprehension for the participants (see Section~\ref{discussion}), as the upward process flow represented an atypical reading direction. Generally, more directions for process flows are inherent in complex process models (e.g., backwards in loops) and, hence, the investigation of other process model mappings (e.g., process flow from right to left) and their influence on performance measures (e.g., eye movements) as well as on the deployment of visual routines may be subject for future work. Thereby, a more precise examination of the made observation (e.g., saturation effect, attention deployed more effectively in the downward mapping) needs to be performed in future studies to be able to fathom their impact. Since process models from practice do not only consist of sequences, but a composition of different modeling constructs (e.g., decision and parallelism), the validation of the model in a more complex process model will enhance our general understanding in process model literacy. Moreover, the process models used in the study documented no semantic information (i.e., abstract element labels; see Section~\ref{material}) and, hence, cognitive capacities in the working memory were lower, compared to labels describing concrete semantic information \cite{mendling2010activity}. In addition to the latter, as known from literature \cite{ceja2020capacity}, cognitive as well as visual capacities are limited to a few elements.  Accordingly, it should be evaluated whether visual routines are exhibited differently (e.g., more frequent revisit of the gaze on an activity) in semantically documented process models, juxtaposed to abstract ones. Because of visual capacity limitations, the visual memory is confronted with challenges during perceptual processing of the information in a process model. More specifically, the quality dimensions within a model (i.e., syntactic, semantic, pragmatic) must be identified, correctly separated, and processed accordingly. Hence, the processing of such information affects the construction of visual routines. Therefore, the identification of precise visual capacity limits during process model comprehension and resulting effects on the construction of such a routine represents another promising research direction. Additionally, this direction may include the evaluation of approaches in order to address these limitations (e.g., application of modularization \cite{winter2021measuring}). In this context, as studied in prior research \cite{michal2017visual}, congruency effects (e.g., congruent vs. incongruent; anchors on comprehension-relevant activities) and their influence on visual routines for finding respective information should be examined in future work. Related research may address the issue of a serial or parallel deployment of elemental operations for the processing of information presented in process models. Further, the determination of an appropriate structure (e.g., multidimensional array) for the representation of visual routines can contribute towards a better understanding of such routines during process model comprehension and should be studied further. Thereupon, such a structure allows for the identification of rudimentary visual routines and related operations in general. The role of the mental model created during process model comprehension, constitutes another promising aspect for future studies. Moreover, the exploration of person-related characteristics in process model literacy and the relation towards visual routines should be pursued as well. In more detail, for example, does expertise in working with process models or general demographic aspects (e.g., gender, age, profession) influence the deployment of visual routines? To double-check the obtained insights, the examination of the association of visual routines as well as the identified gaze patterns should be verified using other process modeling notations (e.g., Event-driven Process Chains, UML Activity Diagram) and paradigms (e.g., data-centric modeling approach \cite{steinau2019dalec}). To summarize, if we know how the visual system (i.e., cognitive control) processes visual information contained in a process model by means of visual routines and elemental operations, respective measures can be introduced supporting this approach.

\section{Conclusion}
\label{conclusion}

This paper presented the first insights from an exploratory eye tracking study, in which visual routines in the context of process model literacy were contemplated. More specifically, a total of n = 29 participants comprehended n = 18 BPMN 2.0 process models reflecting diverse mappings (i.e., straight, upward, downward) as well as complexity levels (i.e., easy, medium, hard). In general, the results indicated a significant increase in the performance measures with a rising level of model complexity. Further, process models depicted with an upward mapping appeared to be more difficult to comprehend, juxtaposed to the other two mapping types. Thereby, the attention of the participants was deployed more effectively in the comprehension of models reflecting a downward mapping. In addition, the study demonstrated that even less complex process models confront participants with challenges regarding proper model comprehension. Further differences in the attention (e.g., eye movements reflecting specific scanning patterns) during process model comprehension were identified. As a result, three common gaze patterns (i.e., orientation, comprehension, congruence) applied during process model comprehension were unraveled. On the basis of these patterns, a model summarizing visual routines and associated elemental operations for process model comprehension was presented. The model indicates how the visual system works in extracting and processing of visual information contained in a process model. As the first explorative work investigating visual routines in process model literacy, the paper makes a contribution to existing research as well as to our conceptual framework, which incorporates measurement methods and theories from cognitive neuroscience and psychology in order to foster our understanding in process model literacy \cite{zimoch1}. While the first results look promising and provide novel insights in this context, however, the conclusions have to be regarded as preliminary. Further research (e.g., addressing confronted limitations) is needed to obtain more fine-grained results allowing for a generalization. Research on visual routines (i.e., information processing from process models) allows for a more sophisticated approach to study the factors influencing process model comprehension (e.g., better assistant for a model reader). Therefore, the continuation of our research regarding visual routines in process model literacy will be the subject of future work.

\bibliographystyle{ieeetran} 
\bibliography{Winter21}

% that's all folks
\end{document}